%% file: Guillon2017_arxiv.tex
\documentclass[1p]{elsarticle}

\usepackage{natbib}
\usepackage{lineno}
\usepackage[breaklinks=true]{hyperref}
\usepackage{xr}
\usepackage{amsmath}
\usepackage{amssymb}
\usepackage{booktabs}
\usepackage{mathtools}
\usepackage{multirow}
\usepackage{graphicx}
\usepackage{color}
\usepackage{soul}
\usepackage{setspace}
\usepackage[author={Jeremy Guillon}]{pdfcomment}
\usepackage{nameref}

\setcitestyle{aysep={}}
\newcommand{\bigcdot}{\raisebox{-0.25ex}{\scalebox{1.2}{$\cdot$}}}
\newcommand{\Hz}{\ensuremath\text{Hz}}
\definecolor{yellow}{rgb}{1,1,0}

\modulolinenumbers[1]
\biboptions{round,sort&compress}
\begin{document}

\begin{frontmatter}

	%%%%%%%%%%%%%%%%%%%%%%%%%%%%%%%%%%%%%%%%%%%%%%%%%%%%%%%%%%
	%%   TITLE                                              %%
	%%%%%%%%%%%%%%%%%%%%%%%%%%%%%%%%%%%%%%%%%%%%%%%%%%%%%%%%%%

	%\title{Multiplex node participation in multifrequency connectomes: application to Alzheimer's disease}
	% OLD: \title{Multifrequency MEG connectomes In Alzheimer’s Disease: A Multilayer Network Approach}
	% Reduced frequency participation of functional connectivity in Alzheimer's disease.
	% Functional connectivity concentrates in lower frequency bands in AD.
	%\title{Multi-frequency brain network reorganization in Alzheimer's disease}
	%\title{Loss of inter-frequency participation in Alzheimer's disease}
	%\title{Loss of inter-frequency centrality in Alzheimer's disease}
	%\title{Loss of inter-frequency brain "hubs" in Alzheimer's disease} %catalina
	%\title{Loss of inter-frequency brain connectivity in Alzheimer's disease}
	%\title{Loss of multi-frequency brain network centrality in Alzheimer's disease}
	\title{Loss of brain inter-frequency hubs in Alzheimer's disease} %catalina

	\author[ARAMISLABAddress,ICMAddress]{J. Guillon}
	\author[MyBrainAddress]{Y. Attal}
	\author[ARAMISLABAddress,ICMAddress]{O. Colliot}
	\author[DescartesAddress,MemAndCogLabAddress]{V. La Corte}
	\author[IM2AAddress]{B. Dubois}
	\author[ICMAddress]{D. Schwartz}
	\author[ICMAddress]{M. Chavez}
	\author[ARAMISLABAddress,ICMAddress]{F. De Vico Fallani\corref{mycorrespondingauthor}}
	\cortext[mycorrespondingauthor]
	{
		%Institut du cerveau et la moelle, Hopital Pitie-Salpetriere \\
		%75013 Paris, France \\
		Corresponding author. \\
		Email fabrizio.devicofallani@gmail.com \\
	}
	\address[ARAMISLABAddress]{Inria Paris, Aramis project-team, 75013, Paris, France}
	\address[ICMAddress]{CNRS UMR-7225, Sorbonne Universites, UPMC Univ Paris 06, Inserm U-1127, Institut du cerveau et la moelle (ICM), Hopital Pitie-Salpetriere, 75013, Paris, France}
	\address[MyBrainAddress]{myBrain Technologies, Paris, France}
	\address[IM2AAddress]{Department of Neurology, Institut de la Mémoire et de la Maladie d’Alzheimer - IM2A, Paris, France}
	\address[DescartesAddress]{Institute of Psychology, University Paris Descartes, Sorbonne Paris Cite, France}
	\address[MemAndCogLabAddress]{INSERM UMR 894, Center of Psychiatry and Neurosciences, Memory and Cognition Laboratory, Paris, France}

	%%%%%%%%%%%%%%%%%%%%%%%%%%%%%%%%%%%%%%%%%%%%%%%%%%%%%%%%%%
	%%   ABSTRACT                                           %%
	%%%%%%%%%%%%%%%%%%%%%%%%%%%%%%%%%%%%%%%%%%%%%%%%%%%%%%%%%%

	\begin{abstract}
		\input{Abstract.tex}
	\end{abstract}


\end{frontmatter}

%\linenumbers

\newpage

%%%%%%%%%%%%%%%%%%%%%%%%%%%%%%%%%%%%%%%%%%%%%%%%%%%%%%%%%%
%%   INTRODUCTION                                       %%
%%%%%%%%%%%%%%%%%%%%%%%%%%%%%%%%%%%%%%%%%%%%%%%%%%%%%%%%%%

\section{Introduction} \label{sec:introduction}

\input{Introduction.tex}

%%%%%%%%%%%%%%%%%%%%%%%%%%%%%%%%%%%%%%%%%%%%%%%%%%%%%%%%%%
%%   METHODS                                            %%
%%%%%%%%%%%%%%%%%%%%%%%%%%%%%%%%%%%%%%%%%%%%%%%%%%%%%%%%%%

\section{Methods} \label{sec:methods}

\input{Methods.tex}

%%%%%%%%%%%%%%%%%%%%%%%%%%%%%%%%%%%%%%%%%%%%%%%%%%%%%%%%%%
%%   RESULTS                                            %%
%%%%%%%%%%%%%%%%%%%%%%%%%%%%%%%%%%%%%%%%%%%%%%%%%%%%%%%%%%

\section{Results} \label{sec:results}

\input{Results.tex}

%%%%%%%%%%%%%%%%%%%%%%%%%%%%%%%%%%%%%%%%%%%%%%%%%%%%%%%%%%
%%   CONCLUSION                                         %%
%%%%%%%%%%%%%%%%%%%%%%%%%%%%%%%%%%%%%%%%%%%%%%%%%%%%%%%%%%

\section{Discussion} \label{sec:discussion}

\input{Discussion.tex}

\section{Conclusions}
%% Final Conclusion

We proposed a multi-layer network approach to characterize multi-frequency brain networks in Alzheimer's disease.
The obtained results gave new insights into the neural deterioration of Alzheimer's disease by revealing an abnormal loss of inter-frequency centrality in memory-related association areas as well as in the cingulate cortex.
Longitudinal studies, including prodromal mild cognitive impairment subjects, will need to assess the predictive value of this new information as a potential non-invasive biomarker for neurodegenerative diseases.
%%%%%%%%%%%%%%%%%%%%%%%%%%%%%%%%%%%%%%%%%%%%%%%%%%%%%%%%%%
%%   ACKNOWLEDGMENTS                                    %%
%%%%%%%%%%%%%%%%%%%%%%%%%%%%%%%%%%%%%%%%%%%%%%%%%%%%%%%%%%

\section*{Acknowledgments}

%\small{
We are grateful to F. Battiston for his useful comments and suggestions. This work has been partially supported by the program ``Investissements d'avenir'' ANR-10-IAIHU-06. FD acknowledges support from the ``Agence Nationale de la Recherche'' through contract number ANR-15-NEUC-0006-02.
The content is solely the responsibility of the authors and does not necessarily represent the official views of any of the funding agencies.
%}

%%%%%%%%%%%%%%%%%%%%%%%%%%%%%%%%%%%%%%%%%%%%%%%%%%%%%%%%%%
%%   BIBLIOGRAPHY                                       %%
%%%%%%%%%%%%%%%%%%%%%%%%%%%%%%%%%%%%%%%%%%%%%%%%%%%%%%%%%%

%\newpage
\bibliography{Guillon2016.bib}

%%%%%%%%%%%%%%%%%%%%%%%%%%%%%%%%%%%%%%%%%%%%%%%%%%%%%%%%%%
%%   FIGURES                                            %%
%%%%%%%%%%%%%%%%%%%%%%%%%%%%%%%%%%%%%%%%%%%%%%%%%%%%%%%%%%

\newpage
\section*{Figures and tables}

\input{Figures.tex}

%%%%%%%%%%%%%%%%%%%%%%%%%%%%%%%%%%%%%%%%%%%%%%%%%%%%%%%%%%
%%   SUPPLEMENTARY MATERIAL                             %%
%%%%%%%%%%%%%%%%%%%%%%%%%%%%%%%%%%%%%%%%%%%%%%%%%%%%%%%%%%

%\appendix

\newpage
\section*{Supplementary Material}

\input{TextS.tex}

\input{SupplementaryMaterial.tex}

\end{document}

%% file: Abstract.tex
% !TEX root = Guillon2017_arxiv.tex
%\small{
Alzheimer's disease (AD) causes alterations of brain network structure and function. The latter consists of connectivity changes between oscillatory processes at different frequency channels.
We proposed a multi-layer network approach to analyze multiple-frequency brain networks inferred from magnetoencephalographic recordings during resting-states in AD subjects and age-matched controls. 

Main results showed that brain networks tend to facilitate information propagation across different frequencies, as measured by the multi-participation coefficient ($MPC$).
However, regional connectivity in AD subjects was abnormally distributed across frequency bands as compared to controls, causing significant decreases of $MPC$. This effect was mainly localized in association areas and in the cingulate cortex, which acted, in the healthy group, as a true inter-frequency hub. 

$MPC$ values significantly correlated with memory impairment of AD subjects, as measured by the total recall score. Most predictive regions belonged to components of the default-mode network that are typically affected by atrophy, metabolism disruption and amyloid-$\beta$ deposition. We evaluated the diagnostic power of the $MPC$ and we showed that it led to increased classification accuracy ($78.39\%$) and sensitivity ($91.11\%$).

These findings shed new light on the brain functional alterations underlying AD and provide analytical tools for identifying multi-frequency neural mechanisms of brain diseases.
%}

%diversity of intermodular connections of individual nodes

%inter-modular inter-frequency hubs facilitates the tranistion of information from one frequency to another one

%% file: Introduction.tex
% !TEX root = Guillon2016.tex

% Intro - generic
Recent advances in network science has allowed new insights in the brain organization from a system perspective.
Characterizing brain networks, or connectomes, estimated from neuroimaging data as graphs of connected nodes has not only pointed out important network features of brain functioning - such as smallworldness, modularity, and regional centrality - but it has also led to the development of biomarkers quantifying reorganizational mechanisms of disease \citep{stam_modern_2014}.
% Alzheimer - the clinical context
Among others, Alzheimer's disease (AD), which causes progressive cognitive and functional impairment, has received great attention by the network neuroscience community \citep{stam_modern_2014,tijms_alzheimers_2013,stam_use_2010}.
AD is histopathologically defined by the presence of amyloid-$\beta$ plaques and tau-related neurofibrillary tangles, which cause loss of neurons and synapses in the cerebral cortex and in certain subcortical regions \citep{tijms_alzheimers_2013}. This loss results in gross atrophy of the affected regions, including degeneration in the temporal and parietal lobe, and parts of the frontal cortex and cingulate gyrus \citep{wenk_neuropathologic_2003}.

Structural brain networks, whose connections correspond to inter-regional axonal pathways are therefore directly affected by AD because of connectivity disruption in several areas including cingulate cortices and hippocampus \citep{rose_loss_2000,zhou_abnormal_2008}.
A decreased number of fiber connections eventually lead to a number of network changes on multiple topological scales.
At larger scales, AD brain networks estimated from diffusion tensor imaging (DTI) showed increased characteristic path length as compared to healthy subjects leading to a global loss of network smallworldness \citep{lo_diffusion_2010,tijms_alzheimers_2013}.
Similar topological alterations have been also documented in resting-state brain networks estimated from functional magnetic resonance imaging (fMRI) \citep{sanz-arigita_loss_2010-1}, as well as from magneto/electroencephalographic (M/EEG) signals, the latter ones often reported within the \textit{alpha} frequency range ($8-13$ Hz) which is typically affected in AD \citep{stam_graph_2009,de_haan_functional_2009,miraglia_searching_2017}.
On smaller topological scales, structural brain network studies have demonstrated a loss of connector hubs in temporal and parietal areas that correlates with cognitive decline \citep{bassett_dynamic_2011,tijms_alzheimers_2013,crossley_hubs_2014}.
In addition, higher-order association regions appear to be affected in functional brain networks inferred from fMRI \citep{buckner_cortical_2009,tijms_alzheimers_2013} and MEG signals, the latter showing a characteristic loss of parietal hubs in higher ($>14$ Hz) frequency ranges \citep{de_haan_disrupted_2012,engels_declining_2015}.

% Despite graph analysis of brain networks has advanced our understanding of the organizational mechanisms underlying human cognition and disease, a certain number of issues still remain to be addressed \citep{de_vico_fallani_graph_2014,bullmore_complex_2009}.
Graph analysis of brain networks has advanced our understanding of the organizational mechanisms underlying human cognition and disease, but a certain number of issues still remain to be addressed \citep{de_vico_fallani_graph_2014,bullmore_complex_2009}.
For example,  conventional approaches analyze separately brain networks obtained at different frequency bands, or in some cases, they simply focus on specific frequencies, thus neglecting possible insights of other spectral contents on brain functioning \citep{de_vico_fallani_graph_2014}.
However, several studies have hypothesized and reported signal interaction or modulations between different frequency bands that are supportive of cognitive functions such as memory formation \citep{canolty_functional_2010, jirsa_cross-frequency_2013,brookes_multi-layer_2016-1}.
Moreover, recent evidence shows that neurodegenerative processes in AD do alter functional connectivity in different frequency bands \citep{fraga_characterizing_2013,engels_declining_2015, blinowska_functional_2016}.
How to characterize this multiple information from a network perspective still remains poorly explored.
Here, we proposed a multi-layer network approach to study multi-frequency connectomes as networks of interconnected layers, containing the connectivity maps extracted from different bands.
Multi-layer network theory has been previously used to synthesize MEG connectomes from a whole population \citep{ghanbari_functionally_2014}, characterize temporal changes in dynamic fMRI brain networks \citep{bassett_dynamic_2011}, and integrating structural information from multimodal imaging (fMRI, DTI) \citep{simas_algebraic_2015, battiston_multilayer_2016}.
Its applicability to multi-frequency brain networks has been recently illustrated in fMRI connectomes for which, however, the frequency ranges of interest remains quite limited \citep{de_domenico_mapping_2016}.

We focused on source-reconstructed MEG connectomes, characterized by rich frequency dynamics, that were obtained from a group of AD and control subjects in eyes-closed resting-state condition.
% We focused on functional brain connectivity networks computed from source-reconstructed MEG signals, characterized by rich frequency dynamics, that were obtained from a group of AD and control subjects in eyes-closed resting-state condition.
We hypothesized that the atrophy process in AD would lead to an altered distribution of regional connectivity across different frequency bands and we used the multiplex participation coefficient to quantify this effect both at global and local scale \citep{battiston_structural_2014}.
We evaluated the obtained results, which provide a novel view of the brain reorganization in AD, with respect to standard approaches based on single-layer network analysis and flattening schemes \citep{de_domenico_mathematical_2013}.
Finally, we tested the diagnostic power of the measured brain network features to discriminate AD patients and healthy subjects.

%% file: Methods.tex
% !TEX root = Guillon2017_arxiv.tex

\subsection{Experimental design and data pre-processing}

The study involved 25 Alzheimer's diseased (AD) patients (13 women) and 25 healthy age-matched control (HC) subjects (18 women).
All participants underwent the Mini-Mental State Examination (MMSE) for global cognition \cite{folstein_mini-mental_1975} and the Free and Cued Selective Reminding Test (FCSRT) for verbal episodic memory \citep{buschke_cued_1984, grober_screening_1988, pillon_explicit_1993}. Specifically, we considered the Total Recall (TR) score - given by the sum of the free and cued recall scores - which has been demonstrated to be highly predictive of AD \citep{sarazin_amnestic_2007}.

Inclusion criteria for all participants were: \textit{i)} age between 50 and 90; \textit{ii)} absence of general evolutive pathology; \textit{iii)} no previous history of psychiatric diseases; \textit{iv) }no contraindication to MRI examination; \textit{v)} French as a mother tongue.
Specific criteria for AD patients were: \textit{i)} clinical diagnosis of Alzheimer's disease;\textit{ ii)} Mini-Mental State Examination (MMSE) score greater or equal to $18$.
Magnetic resonance imaging (MRI) acquisitions were obtained using a 3T system (Siemens Trio, 32-channel system, with a 12-channel head coil). The MRI examination included a 3D T1-weighted volumetric magnetization-prepared rapid-gradient echo (MPRAGE) sequence with 1mm isotropic resolution and the following parameters: repetition time (TR)=2300 ms, echo time (TE)=4.18ms, inversion time (TI)=900 ms, matrix=256x256. This sequence provided a high contrast-to-noise ratio and enabled excellent segmentation of high grey/white matter.

The magnetoencephalography (MEG) experimental protocol consisted in a resting-state with eyes-closed (EC). Subjects seated comfortably in a dimly lit electromagnetically and acoustically shielded room and were asked to relax and fix a central point on the screen.
MEG signals were collected using a whole-head MEG system with $102$ magnetometers and $204$ planar gradiometers (Elekta Neuromag TRIUX MEG system) at a sampling rate of $1\,000$ Hz and on-line low-pass filtered at $330$ Hz.
The ground electrode was located on the right shoulder blade. An electrocardiogram (EKG) Ag/AgCl electrodes was placed on the left abdomen for artifacts correction and a vertical electrooculogram (EOG) was simultaneously recorded.
% Horizontal and vertical eye position as well as pupil diameter were monitored using an eye tracker (EyeLink 1000, SR research) and recorded together with MEG and EKG data. % => Useless since eyes are closed.
Four small coils were attached to the participant in order to monitor head position and to provide co-registration with the anatomical MRI. The physical landmarks (the nasion, the left and right pre-auricular points) were digitized using a Polhemus Fastrak digitizer (Polhemus, Colchester, VT).

We recorded three consecutive epochs of approximately $2$ minutes each. All subjects gave written informed consent for participation in the study, which was approved by the local ethics committee of the Pitie-Salpetriere Hospital. Signal space separation was performed using MaxFilter \citep{taulu_spatiotemporal_2006} to remove external noise. We used in-house software to remove  cardiac and ocular blink artifacts from MEG signals by means of principal component analysis.
We visually inspected the preprocessed MEG signals in order to remove epochs that still presented spurious contamination.
At the end of the process, we obtained a coherent dataset consisting of three clean preprocessed epochs for each subject.

\subsection{Source reconstruction, power spectra and brain connectivity} \label{subsec:brain_connectivity}

We reconstructed the MEG activity on the cortical surface by using a source imaging technique \citep{he_brain_1999,baillet_evaluation_2001}.
We used the FreeSurfer 5.3 software (surfer.nmr.mgh.harvard.edu) to perform skull stripping and segment grey/white matter from the 3D T1-weighted images of each single subject \citep{fischl_whole_2002, fischl_sequence-independent_2004}. Cortical surfaces were then modeled with approximately $20000$ equivalent current dipoles (i.e., the vertices of the cortical meshes).
We used the Brainstorm software \citep{tadel_brainstorm:_2011} to solve the linear inverse problem though the wMNE (weighted Minimum Norm Estimate) algorithm with overlapping spheres \citep{lin_assessing_2006}. Both magnetometer and gradiometer, whose position has been registered on the T1 image using the digitized head points, were used to localize the activity over the cortical surface.
The reconstructed time series were then extracted from $148$ regions of interest (ROIs) defined by the Destrieux atlas \citep{destrieux_automatic_2010-1}.
% We chose this Atlas for its number of ROIs that is computationally-wise adapted for this application and close to the number of MEG sensors, because it is provided by the FreeSurfer software, because it is a cortical-only atlas and because the regions have similar size hence a similar number of vertices on the cortical mesh.

We computed the power spectral density (PSD) of the ROI signals by means of the Welch's method; we chose a $2$ seconds sliding Hanning window, with a $25\%$ overlap. The number of FFT points was set to $500$ for a frequency resolution of $0.5 \Hz$.
% NOTE: Finding reference for spectral coherence is not useful.
We estimated functional connectivity by calculating the spectral coherence between each pair of ROI signals \citep{carter_coherence_1987}. 
%For a given frequency $f$, the spectral coherence for the channels pair $(i,j)$ can be computed as follow:
%\begin{equation}
	%Coh_{ij}(f) = \frac{ \abs{S_{ij}(f)} }{ \sqrt{S_{ii}(f)S_{jj}(f)} }
	%\label{eq:coherence}
%\end{equation}
%where $S_{ij}(f)$ is the cross-spectrum of two time series $x_i(t)$ and $x_j(t)$ of ROI $i$ and $j$ respectively. We used the same FFT parameters as for the PSD.
As a result, we obtained for each subject and epoch, a connectivity matrix of size $148 \times 148$ where the $(i,j)$ entry contains the value of the spectral coherence between the signals of the ROI $i$ and $j$ at a frequency $f$.

We then averaged the connectivity matrices within the following characteristic frequency bands \citep{stam_generalized_2002,babiloni_abnormal_2004}: $\delta$ (2-4 Hz); $\theta$ (4-8 Hz); $\alpha=\alpha_{1}$ (8-10.5 Hz) and $\alpha_{2}$ (10.5-13 Hz); $\beta=\beta_{1}$ (13-20 Hz) and $\beta_{2}$ (20-30 Hz); $\gamma$ (30-45 Hz).
We further averaged the resulting connectivity matrices across epochs to obtain our raw individual brain networks whose nodes were the ROIs ($n = 148$) and links, or edges, were the spectral coherence values.

\subsection{Single-layer network analysis} \label{subsec:singlelayer}

In order to cancel the weakest noisy connections, we thresholded the values in the connectivity matrices and retained the same number of links in each brain network at every frequency band, or layer.
We considered six representative connection density thresholds corresponding to an average node degree $k=\{1, 3, 6, 12, 24, 48\}$. These values cover the density range $[0.007, 0.327]$ which contains the typical density values used in complex brain network analysis \citep{bullmore_complex_2009,rubinov_complex_2010,de_vico_fallani_graph_2014}.
The resulting sparse brain networks, or graphs,  were represented by adjacency matrices $A$, where the $a_{ij}$ entry indicates the presence or absence of a link between nodes $i$ and $j$.

\subsubsection{Participation coefficient} \label{subsec:pc}

Given a network partition, the local participation coefficient ($PC_i$) of a node $i$ measures how evenly it is connected to the different clusters, or modules of the network \citep{guimera_cartography_2005}.
Nodes with high participation coefficients are considered central hubs as they allow for the information exchange among different modules.
The global participation coefficient $PC$ of a network at layer $\lambda$  is then given by the average of the $PC_i$ values:
\begin{equation}
	PC^{[\lambda]} 	= \frac{1}{n} \sum_{i=1}^{N} PC_i^{[\lambda]}
	=  \frac{1}{n} \sum_{i=1}^{N} \Bigg[ 1-\sum_{m=1}^{M^{[\lambda]}} \Bigg( \frac{k_{i,m}^{[\lambda]}}{k_i^{[\lambda]}} \Bigg)^2 \Bigg] \text{,}
	\label{eq:pc}
\end{equation}
where $k_{i,m}^{[\lambda]}$ is the number of weighted links from the node $i$ to the nodes of the module $m$ of the layer $\lambda$. By construction, $PC$ ranges from $0$ to $1$.
Here, the partition of the networks into modules was obtained by maximizing the modularity function as defined by \cite{newman_finding_2006}.

\subsubsection{Flattened networks} \label{subsec:flattening}

We also computed the participation coefficients for brain networks obtained by flattening the frequency layers into a single \textit{overlapping}  or \textit{aggregated} network \citep{battiston_structural_2014}.
In an overlapping network, the weight of an edge $o_{ij}$ corresponds to the number of times that the nodes $i$ and $j$ are connected across layers:
\begin{equation}
	o_{ij} = \sum_{\lambda} a_{ij}^{[\lambda]} \text{,}
	\label{eq:overlapping}
\end{equation}

In an aggregated network, the existence of an edge indicates that nodes $i$ and $j$ are connected in at least one layer:
\begin{equation}
	a_{ij} = \left\{
	\begin{array}{ll}
		1 & \text{if } \exists{\lambda} : a_{ij}^{[\lambda]} \ne 0 \\
		0 & \text{otherwise}                                       \\
	\end{array}
	\right. \text{,}
	\label{eq:aggregated}
\end{equation}

Notice that, by construction, flattened networks do not preserve the original connection density of the single layer networks.

\subsection{Multi-layer network analysis} \label{subsec:multiplex_networks}
% TODO: Add reference for multiplex definition
We adopted a multi-layer network approach to integrate the information from brain networks at different frequency bands, while preserving their original structure.
We built for each subject a multiplex network (Fig. \ref{fig:multiplex}a,b) where different layers correspond to different frequency bands and each node in one layer is virtually connected to all its counterparts in all the other layers.

Without loss of generality, if we consider the standard neurophysiological frequency bands, the resulting supra-adjacency matrix $\mathcal{A}$ is given by the following intra-layer of adjacency matrices on the main diagonal:
\begin{equation}
	\mathcal{A} = \{ A^{[\delta]}, A^{[\theta]}, A^{[\alpha]}, A^{[\beta]}, A^{[\gamma]} \}\text{,}
	\label{eq:supraadjacency}
\end{equation}
where $A^{[\lambda]}$ is adjacency matrix of the frequency layer $\lambda$. By construction, the inter-layer adjacency matrices of multiplexes are intrinsically defined as identity matrices.

\subsubsection{Multi-participation coefficient} \label{subsec:mpc}
We considered the multi-layer version of the local participation coefficient $MPC_i$ to measure how evenly a node $i$ is connected to the different layers of the multiplex \citep{battiston_structural_2014}.
This way, nodes with high multi-participation coefficients are considered central hubs as they would allow for a better information exchange among different layers.
The global multi-participation coefficient is then given by the average of the $MPC_i$ values:

\begin{equation}
	MPC = \frac{1}{n} \sum_{i=1}^{N} MPC_i
	= \frac{1}{n} \sum_{i=1}^{N} \frac{M}{M-1} \Bigg[ 1-\sum_{\lambda} \Bigg( NLP_i^{[\lambda]} \Bigg)^2 \Bigg] \text{,}
	\label{eq:mpc}
\end{equation}
where $NLP_i^{[\lambda]} = k_i^{[\lambda]} / o_i$, stands for \textit{node-degree layer proportion}, which measures the percentage of the total number of links (i.e. in all layers) of node $i$ that are in layer $\lambda$.
By construction, if nodes tend to concentrate their connectivity in one layer, the global multi-participation coefficient tends to $0$; on the contrary, if nodes tend to have the same number of connections in every layer, the $MPC$ value tends to  $1$ (Fig. \ref{fig:multiplex}c). Hence, a node with a high $MPC$ has the potential to facilitate communication across layers.
The Matlab code for the computation of the $MPC$ is freely available at \href{https://github.com/devuci/BNT}{https://github.com/devuci/BNT}.

We also used the standard coefficient of variation $CV_i$ to measure the dispersion of the degree of a node $i$ across layers. A global coefficient of variation $CV$ is then obtained by averaging the $CV_i$ values across all the nodes  (\nameref{subsec:TextS}).

\subsection{Statistical analysis}

We first analyzed network features on global topological scales in order to detect statistical differences between AD and HC subjects at the whole system level.
Only for those conditions (e.g., frequency bands) that resulted significantly different on the global scale, we also assessed possible group-differences on the local topological scale of single nodes.
This hierarchical approach allowed us to associate brain network differences on multiple topological scales \citep{de_vico_fallani_interhemispheric_2016}.
For global network features, we used a non-parametric permutation t-test to assess statistical differences between groups, with a significance level of $0.05$. For local network features, we applied a correction for multiple comparisons by computing the rough false discovery rate (FDR) \citep{benjamini_controlling_1995,zar_biostatistical_1999}.
In both cases, surrogate data were generated by randomly exchanging the group labels $10\,000$ times.

%hierarchical analysis
To test the ability of the significant brain network properties to predict the cognitive/memory impairment of AD patients, we used the non-parametric Spearman's correlation coefficient $R$. We set a significance level of $0.05$ for the correlation of global network features, with a FDR correction in the case of multiple comparisons (local features).

\subsection{Classification}

We used a classification approach to evaluate the discriminating power of the local brain network properties which resulted significantly different in the AD and HC group.
Because we did not know in advance which were the most discriminating features, we tested different combinations. In particular, for each local network property, we first ranked the respective ROIs according to the $p$-values returned by the between-group statistical analysis (see previous section).
For each subject $s$, we then tested different feature vectors obtained by concatenating, one-by-one, the values of the network features extracted from the ranked ROIs.
The generic feature vector $c_s$ reads:
\begin{equation}
	c_s =  [g_1, ... ,g_k]
	\label{eq:caractestics}
\end{equation}
where $g_k$ is a generic local network feature and $k$ is a rank that ranges from $1$ (the most significant ROI) to the total number of significant ROIs.
When different network properties were considered (e.g., $PC$ and $MPC$), we concatenated the respective $c_s$ feature vectors allowing for all the possible combinations.
% TODO: Possibly explain better

% Classification algorithm
To quantify the separation between the feature vectors of AD and HC subjects, we used a Mahalanobis distance classifier.
We applied a repeated $5$-fold cross-validation procedure where we randomly split the entire dataset into a training set ($80\%$)  and a testing test ($20\%$). This procedure was eventually iterated $10\,000$ times in order to obtain more accurate classification rates.
To assess the classification performance we computed the sensitivity ($Sens$), specificity ($Spec$) and accuracy ($Acc$), defined respectively as the percentage of AD subjects correctly classified as AD, the percentage of HC subjects classified as HC and the total percentage of subjects (AD and HC) properly classified.
We also computed the receiver operating characteristic (ROC) curve and its area under the curve ($AUC$) \citep{hastie_elements_2009}.

%% file: Results.tex
% !TEX root = Guillon2017_arxiv.tex

Power analysis of source-reconstructed MEG signals confirmed the characteristic changes in the oscillatory activity of AD subjects compared to HC subjects (\autoref{fig:psd}a) \citep{babiloni_mapping_2004,jeong_eeg_2004,dauwels_diagnosis_2010,wang_power_2015}. Significant \textit{alpha} power decreases were more evident in the parietal and occipital regions ($Z < -2.58$), while significant \textit{delta} power increases ($Z > 2.58$) were more localized in the frontal regions of the cortex (\autoref{fig:psd}b).

\subsection{Reduced \textit{gamma} inter-modular connectivity} \label{subsec:decreased_pc_in_gamma}

As expected the value of the connection density threshold had an impact on the network differences between groups. For the sake of simplicity, we selected the first threshold for which we could observe a significant group difference for both single- and multi-layer analysis. The obtained results determined the choice of a representative threshold, common to all the brain networks, corresponding to an average node degree $k=12$ (\autoref{fig:thresholding}).

We first evaluated the results from the single-layer analysis. By inspecting the global participation coefficient $PC$, we reported in the $gamma$ band a significant decrease of inter-modular connectivity in AD as compared to HC ($Z=-2.50$, $p=0.017$; \autoref{fig:participation}a inset).
This behavior was locally identified in association ROIs including temporal and parietal areas ($p < 0.05$, FDR corrected; \autoref{fig:participation}a; \autoref{tab:local_participation}).
No other significant differences were reported in other frequency bands or in flattened brain networks (\autoref{fig:thresholding}).

% Global participation coefficient 100.000-permutations test for the seven frequency bands, MATLAB Output :
%   pValue =   0.5445    0.6551    0.5337    0.9684    0.9302    0.2337    0.0160
%   zScore =  -0.6061   -0.4494    0.6286   -0.0402   -0.0858   -1.2121   -2.4906

\subsection{Disrupted inter-frequency hub centrality}

Then we assessed the results from the multi-layer analysis. Both AD and HC subjects exhibited high global multi-participation coefficients ($MPC>0.9$), suggesting a general propensity of brain regions to promote interactions across frequency bands.
However, such tendency was significantly lower in AD than HC subjects ($Z=-2.24$, $p=0.028$; \autoref{fig:participation}b inset).
This loss of inter-frequency centrality was prevalent in association ROIs including temporal, parietal and cingulate areas, and with a minor extent in motor areas ($p<0.05$, FDR corrected; \autoref{fig:participation}b; \autoref{tab:local_participation}).

Among those regions, the right cingulate cortex was classified as the main inter-frequency hub as revealed by the spatial distribution of the top $25\%$ $MPC$ values in the HC group (\autoref{fig:mpc}a).
%put in this picture the two hemisherepes 25% sectral coherence and the corresponding box plot; a thris panel would be the nlp distribution
% NLP only for cingulate (hub)
In HC subjects the connectivity of this region across bands, as measured by the node degree layer proportion $NLP$, was relatively stable (Kruskal-Wallis test, $\chi^2=10.79$, $p=0.095$), while it was significantly altered in AD subjects (Kruskall-Wallis test, $\chi^2=14.98$, $p=0.020$).
In particular, the AD group exhibited a remarkably reduced $alpha_2$ connectivity and increased $theta$ connectivity (\autoref{fig:mpc}b). Similar results were also reported for the left cingulate cortex (AD: $\chi^2=11.89$, $p=0.064$; HC: $\chi^2=6.98$, $p=0.323$), although it was not significant in terms of $MPC$ differences (\autoref{fig:participation}b; \autoref{tab:local_participation}).

% TODO: NLP for all significant regions
%The connectivity distribution for these significant regions, as measured by the node degree layer proportion $NLP$, was significantly altered in the AD group (Kruskall-Wallis test, $\chi^2=14.35$, $p=0.026$), while it was relatively stable across bands in the HC group (Kruskal-Wallis test, $\chi^2=7.59$, $p=0.270$).
%AD subjects exhibited a decreasing trend with reduced $beta_2$ and $gamma$ connectivity and increased $theta$ and $alpha_1$ connectivity, while a more constant trend was found in HC subjects (\autoref{fig:participation}c).
%In both populations, the contribution of $delta$ connectivity for these ROIs was remarkably low.

% Tijms2013 correspondances:
% ==========================
% Right IFG = G_front_inf-Operocular R (1 AD)
% Right PreCG = G_precentral R (1 AD, 2 HC)
% Right INS = S_temporal_transverse R (1 AD) ???
% Right CING = S_pericallosal R (3 AD, 3 HC) "The cingulate gyrus is limited from the corpus callosum by the pericallosal sulcus [...]" [Destrieux2010]
% Left SMG = G_pariet_inf-Supramar L (1 AD, 1 HC)
% Left MOG = S_oc_middle_and_Lunatus L (2 AD, 5 HC)
% Left ANG = S_interm_prim-Jensen L (2 AD) "The sulcus intermedius primus divides the inferior parietal lobule into supramarginal (anterior) and angular (posterior) gyri." [Destrieux2010]

\subsection{Diagnostic power of brain network features}

We adopted a classification approach to evaluate the power of the most significant local network properties in determining the state (i.e., healthy or diseased) of each individual subject.
The best results were achieved neither when we considered single-layer features (i.e., $PC^{[\gamma]}_i$) nor when we considered multi-layer features ($MPC_i$) (respectively, first column and row of panels in \autoref{fig:classification}a). Instead, a combination of the two most significant features gave the best classification in terms of accuracy ($Acc=78.39\%$) and area under the curve ($AUC=0.8625$)  (\autoref{fig:classification}a,b).
While the corresponding specificity was not particularly high ($Spec=65.68\%$), the sensitivity was remarkably elevated ($Sens=91.11\%$).

% NOTE: The only misclassified AD has the following charecteristics :
% mmseScore: 28
%       age: 81
%       sex: 'F'
%      rimm: 16
%        ri: 21
%        rl: 22
%        rt: 43
%     react: 81
% She has an abnormally high MMSE for an AD patient. Also her total recall and react score are quite similar to those of HC subjects.

\subsection{Relationship with cognitive and memory impairment}

We finally evaluated the ability of the significant brain network changes to predict the cognitive and memory performance of AD subjects.
We first considered the results from single-layer analysis. We found a significant positive correlation between the global participation coefficient $PC$ in the $gamma$ band and the MMSE score ($R=0.4909$, $p=0.0127$; \autoref{fig:correlations}a).
Then we considered the results from multi-layer analysis. We reported a higher significant positive correlation between the global multi-participation coefficient $MPC$ and the TR score ($R = 0.5547$, $p = 0.0074$; \autoref{fig:correlations}c).
These relationships were locally identified in specific ROIs including parietal, temporal and cingulate areas of the default mode network (DMN) \citep{buckner_brains_2008} ($p < 0.05$, FDR corrected; \autoref{fig:correlations}b,d; \autoref{tab:local_correlation}).

%% file: Discussion.tex
% !TEX root = Guillon2017_arxiv.tex

%% Short Recap

Graph analysis of brain networks have been largely exploited in the study of AD with the aim to extract new predictive diagnostics of disease progression.
Typical approaches in functional neuroimaging, characterized by oscillatory dynamics, analyze brain networks separately at different frequencies thus neglecting the available multivariate spectral information.
Here, we adopted a method to formally take into account the topological information of multi-frequency connectomes obtained from source-reconstructed MEG signals in a group of AD and healthy subjects during EC resting states.

%% Multiplex Results

Main results showed that, while flattening networks of different frequency bands attenuates differences between AD and HC populations, keeping the multiplex nature of MEG connectomes allow to capture higher-order discriminant information.
AD subjects exhibited an aberrant multiplex brain network structure that significantly reduced the global propensity to facilitate information propagation across frequency bands as compared to HC subjects (\autoref{fig:participation}b, inset). This could be in part explained by the higher variability of the individual node degrees across bands (\autoref{fig:coefficient_of_variation}).

% NOTE: High MPCi does not necessarily mean high oi (autoref fig:mpca) but it also seems that having a high number of connections (high oi) and a low MPC is not possible in the case of the brain.

% NOTE: In general, a ROI with a high MPC but with low oi, will have an even higher MPC if its oi increase (for another subject for instance). In clear: for a given i (i.e. a given ROI), the corrcoeff between oi and MPCi is always positive.

% NOTE: I tested different thresholds and with the ImCoh to check if it was not because of the week noisy connections, but the distribution of MPCi values seems to be always the same. Even with an average degree of 1 meaning that the brain always tends to keep connections in multiples frequency bands in the same time. Could it be explained by the fact that coherence is influenced by harmonics?

Such loss of inter-frequency centrality was mostly localized in association areas as well as in the cingulate cortex (\autoref{fig:participation}b; \autoref{tab:local_participation}), which resulted the most important hub promoting interaction across bands in the HC group (\autoref{fig:mpc}a).
Because all these areas are typically affected by AD atrophy \citep{wenk_neuropathologic_2003} we hypothesize that the anatomical withering might have impacted the neural oscillatory mechanisms supporting large-scale brain functional integration. Notably, the significant alteration of the connectivity across bands observed in the cingulate cortex could be ascribed to typical M/EEG connectivity changes observed in AD, such as reduced $alpha$ coherence \citep{stam_magnetoencephalographic_2006,jeong_eeg_2004,dauwels_diagnosis_2010,wang_power_2015} (\autoref{fig:mpc}b).
We also found a significant decrease in the primary motor cortex (right precentral gyrus). While previous studies have identified this specific region as a connector hub in human brain networks \citep{tijms_alzheimers_2013}, its role in AD still needs to be clarified in terms of node centrality's changes with respect to healthy conditions.
%For these affected ROIs the decreased centrality was reflected by fewer interactions with higher sensory rhythms ($>20$ Hz) \citep{basar_review_2013} and more connections to lower attentional ones ($<13$ Hz) \citep{klimesch_EEG_1999} (\autoref{fig:participation}c).

% Single-Layer Results
While flattening network layers represents in general an oversimplification, analyzing single layers can still be a valid approach that is worth of investigation.
Because the $MPC$ is a pure multiplex quantity, we considered the conceptually akin version for single-layer networks, the standard participation coefficient $PC$, which evaluates the tendency of nodes to integrate information from different modules, rather than from different layers \citep{guimera_cartography_2005, battiston_structural_2014}.
AD patients exhibited lower inter-modular connectivity in the \textit{gamma} band with respect to HC subjects (\autoref{fig:participation}a; \autoref{tab:local_participation}) that was localized in association areas including frontal, temporal, and parietal cortices (\autoref{fig:participation}a; \autoref{tab:local_participation}).
Damages to these regions can lead to deficits in attention, recognition and planning \citep{purves_neuroscience_2001}. Our results support the hypothesis that AD could include a disconnection syndrome  \citep{pearson_anatomical_1985,arnold_topographical_1991,catani_rises_2005}.
Furthermore, they are in line with previous findings showing $PC$ decrements in AD, although those declines were more evident in lower frequency bands and therefore ascribed to possible long-range low-frequency connectivity alteration \citep{de_haan_disrupted_2012,tijms_alzheimers_2013}.

%% Conclusion
Put together, our findings indicated that AD alters the global brain network organization through connection disruption in several association regions, which play important roles in sensory processing by integrating information from other cortical regions through high-frequency channels \citep{miltner_coherence_1999-1,buschman_top-down_2007, siegel_neuronal_2008, gregoriou_high-frequency_2009, hipp_oscillatory_2011}.
Notably, we showed that the global loss of inter-modular interactions in the \textit{gamma} band is paralleled by a diffused decrease of inter-frequency centrality.
Future studies, involving recordings of limbic structures and/or stimulation-based techniques, should elucidate whether these two distinct reorganizational processes are truly independent or linked through possible cross-frequency mechanisms which are known to be essential for normal memory formation \citep{canolty_high_2006,axmacher_cross-frequency_2010, goutagny_alterations_2013}.

%% Classification Results

As a confirmation of the complementary information carried out by the multi-layer approach, we reported an increased classification accuracy when combining the local $PC$ and $MPC$ features.
The observed diagnostic power is in line with previous accuracy values obtained with standard graph theoretic approaches (around $80\%$) but exhibits slightly higher sensitivity ($>90\%$), which is often desired to avoid false negatives \citep{li_discriminant_2012, wang_disrupted_2013, wee_enriched_2011, wee_identification_2012, horwitz_functional_2011}.
Other approaches should determine if and to what extent the use of more sophisticated machine learning algorithms, or the inclusion of basic connectivity features \citep{hutchison_network-based_2011, shao_prediction_2012, zhou_hierarchical_2011} and different imaging modalities \citep{dai_discriminative_2012}, can lead to higher classification performance and better diagnosis \citep{tijms_alzheimers_2013}.

%% Correlation With MMSE

Previous works have documented relationships between brain network properties and neuropsychological measurements in AD, suggesting a potential impact for monitoring disease progression and for the development of new therapies
\citep{de_haan_functional_2009,lo_diffusion_2010,sanz-arigita_loss_2010-1,shu_disrupted_2012,stam_small-world_2007,wang_disrupted_2013}.
This is especially true for the standard $PC$ which has exhibited stronger correlations and larger between-group differences \citep{tijms_alzheimers_2013}.
In line with this prediction, we also reported significant correlations between the MMSE cognitive scores and the $PC$ values of the AD patients in the \textit{gamma} band (\autoref{fig:correlations}a).
An even stronger correlation was found, however, for the global $MPC$ values and the TR scores (\autoref{fig:correlations}b, \autoref{tab:local_correlation}).
Recent studies suggest that TR scores could be more specific for AD \citep{grober_free_2010, velayudhan_review_2014} as compared to MMSE scores which could be biased by differences in years of education, lack of sensitivity to progressive changes occurring with AD, as well as fail in detecting impairment caused by focal lesions \citep{tombaugh_mini-mental_1992}.
Locally, the regions whose $MPC$ correlated with TR were part of the default-mode network (DMN) (\autoref{tab:local_correlation}), which is heavily involved in memory formation and retrieval \citep{buckner_brains_2008,sperling_functional_2010}. According to recent hypothesis, these areas are directly affected by atrophy and metabolism disruption, as well as amyloid-$\beta$ deposition \citep{buckner_molecular_2005, greicius_default-mode_2004}.
Put together, our results suggest that AD symptoms related to episodic memory losses could be determined by the lower capacity of strategic DMN association areas to let information flow across different frequency channels.

\subsection*{Methodological considerations}

We estimated brain networks by means of spectral coherence, a connectivity measure widely used in the electrophysiological literature because of its simplicity and relatively intuitive interpretation \citep{srinivasan_eeg_2007}.
While this measure is known to suffer from possible volume conduction effects, recent evidence showed that source reconstruction techniques, like the one we adopted here, could at least mitigate this bias \citep{schoffelen_source_2009} and generate connectivity patterns consistent within and between subjects \citep{colclough_how_2016}.
In a separate analysis, we used the imaginary coherence as a candidate alternative to eliminate volume conduction effects \citep{nolte_identifying_2004}. We demonstrated that while no significant between-group differences could be obtained in terms of $MPC$ (data not shown here), the spatial distribution of the $MPC$ values was very similar to that observed in the brain networks obtained with the spectral coherence, especially for the internal regions along the longitudinal fissure (\autoref{fig:mpc_imcoh}).

Differently from other multiplex network quantities, such as those based on paths and walks \citep{boccaletti_structure_2014}, the $MPC$ has the advantage to not depend on the weights of the inter-layer links which, in general, are difficult to estimate or to assign from empirically obtained biological data. This is especially true in network neuroscience where, so far, the strength of the inter-layer connections is parametric and subject to arbitrariness \citep{de_domenico_mapping_2016} or estimated through measures of cross-frequency coupling \citep{brookes_multi-layer_2016-1} whose biological interpretation remains still to be completely elucidated \citep{jirsa_cross-frequency_2013}.

%% file: Figures.tex
% !TEX root = Guillon2017_arxiv.tex

%\newpage
\begin{figure}[!ht]
	\centering
	\includegraphics[width=\textwidth]{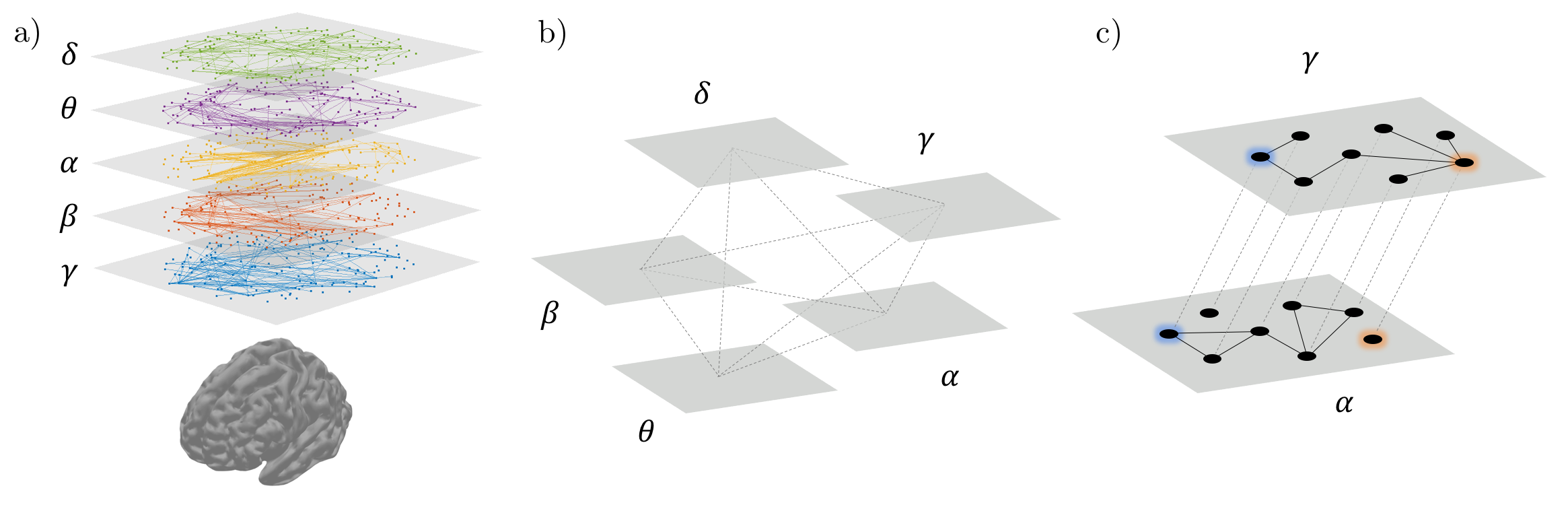}
	\caption{Multi-frequency brain networks.
	Panel a) shows five representative networks extracted from typical frequency bands.
	b) Procedure to construct a multi-frequency network by virtually connecting the homologous brain nodes among frequency layers.
	c) Inter-frequency node centrality. A two-layer multiplex is considered for the sake of simplicity. The blue node acts as an inter-frequency hub (i.e., multi-participation coefficient $MPC=1$) as it allows for a balanced information transfer between layer $\alpha$ and $\gamma$; the red node, who is disconnected in layer $\alpha$, blocks the information flow and has $MPC=0$.}
	\label{fig:multiplex}
\end{figure}

\newpage
\begin{figure}[!ht]
	\centering
	\includegraphics[width=\textwidth]{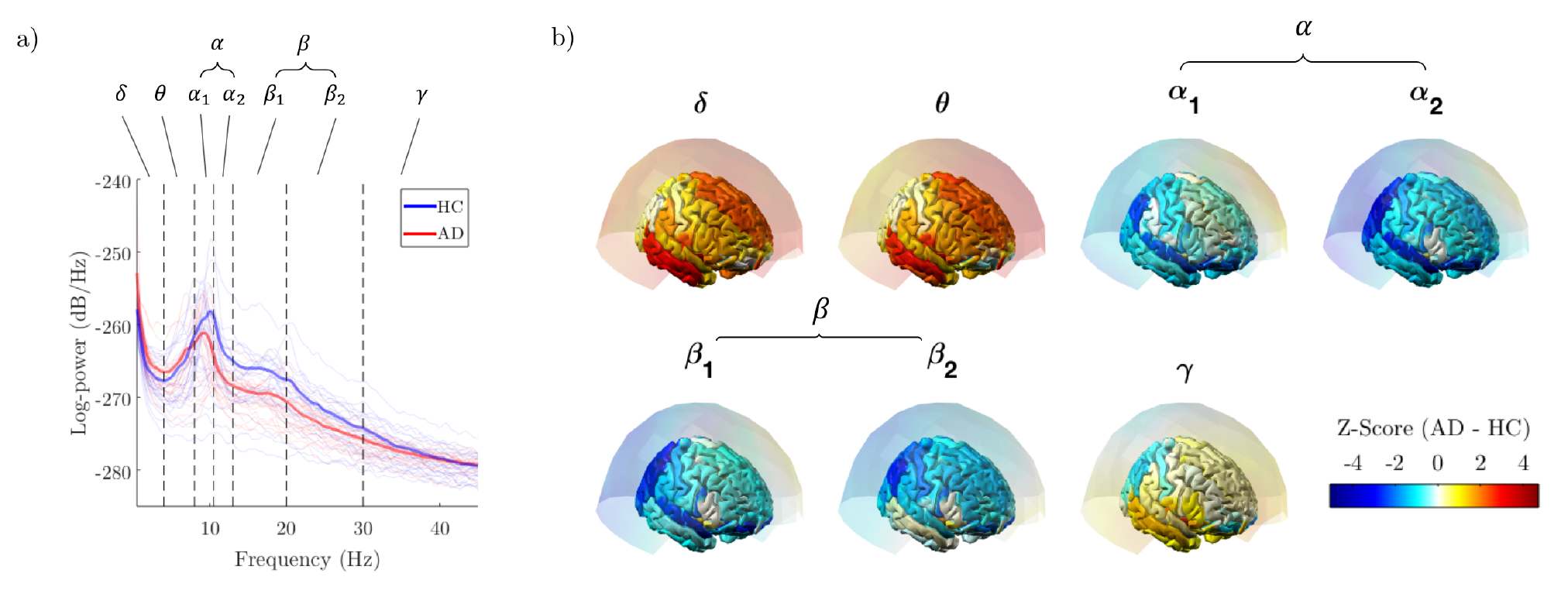}
	\caption{Spectral analysis of MEG signals.
	a) Power spectrum density (PSD) for a representative occipital sensor before source reconstruction. Each line corresponds to a subject. Bold lines show the group-averaged values in the Alzheimer's disease group (AD) and in the healthy control group (HC).
	b) Statistical PSD group differences. Z-scores are obtained using a non-parametric permutation t-test. Results are represented both as sensor and source space.}
	\label{fig:psd}
\end{figure}

\newpage
\begin{figure}[!ht]
	\centering
	\includegraphics[width=10cm]{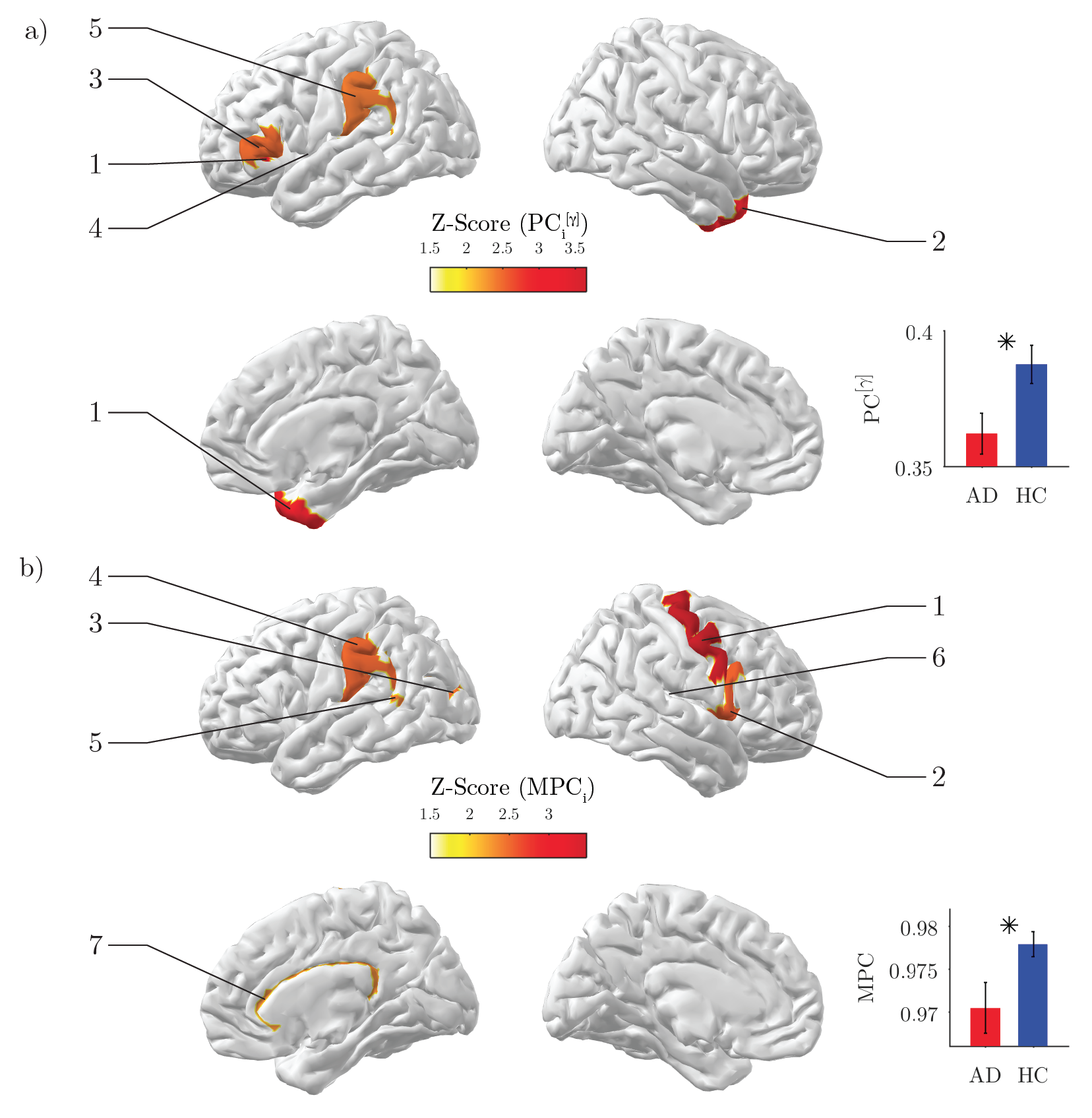}
	\caption{Network analysis of brain connectivity.
	a) Inter-modular centrality. Statistical brain maps of group differences for local participation coefficients $PC_i$ in the $gamma$ band. Only significant differences are illustrated ($p<0.05$, FDR corrected). The labels same ranks are used as labels. The inset shows the results for the global $PC$; vertical bars stand for group-averaged values while error bars denote standard error means. In both cases, Z-scores are computed using a non-parametric permutation t-test.
	b) Inter-frequency centrality. Statistical brain maps of group differences for local multi-participation coefficients $MPC_i$. The inset shows the results for the global $MPC$; same conventions as in a).}
	% Global participation coefficient 100.000-permutations test for the sevent frequency bands, MATLAB Output :
	%
	%   pValue =
	%
	%     0.5445    0.6551    0.5337    0.9684    0.9302    0.2337    0.0160
	%
	%
	%   zScore =
	%
	%    -0.6061   -0.4494    0.6286   -0.0402   -0.0858   -1.2121   -2.4906
	%
	% Of course results vary accross runs (permutation test)
	\label{fig:participation}
\end{figure}

\newpage
\begin{figure}[!ht]
	\centering
	\includegraphics[width=14cm]{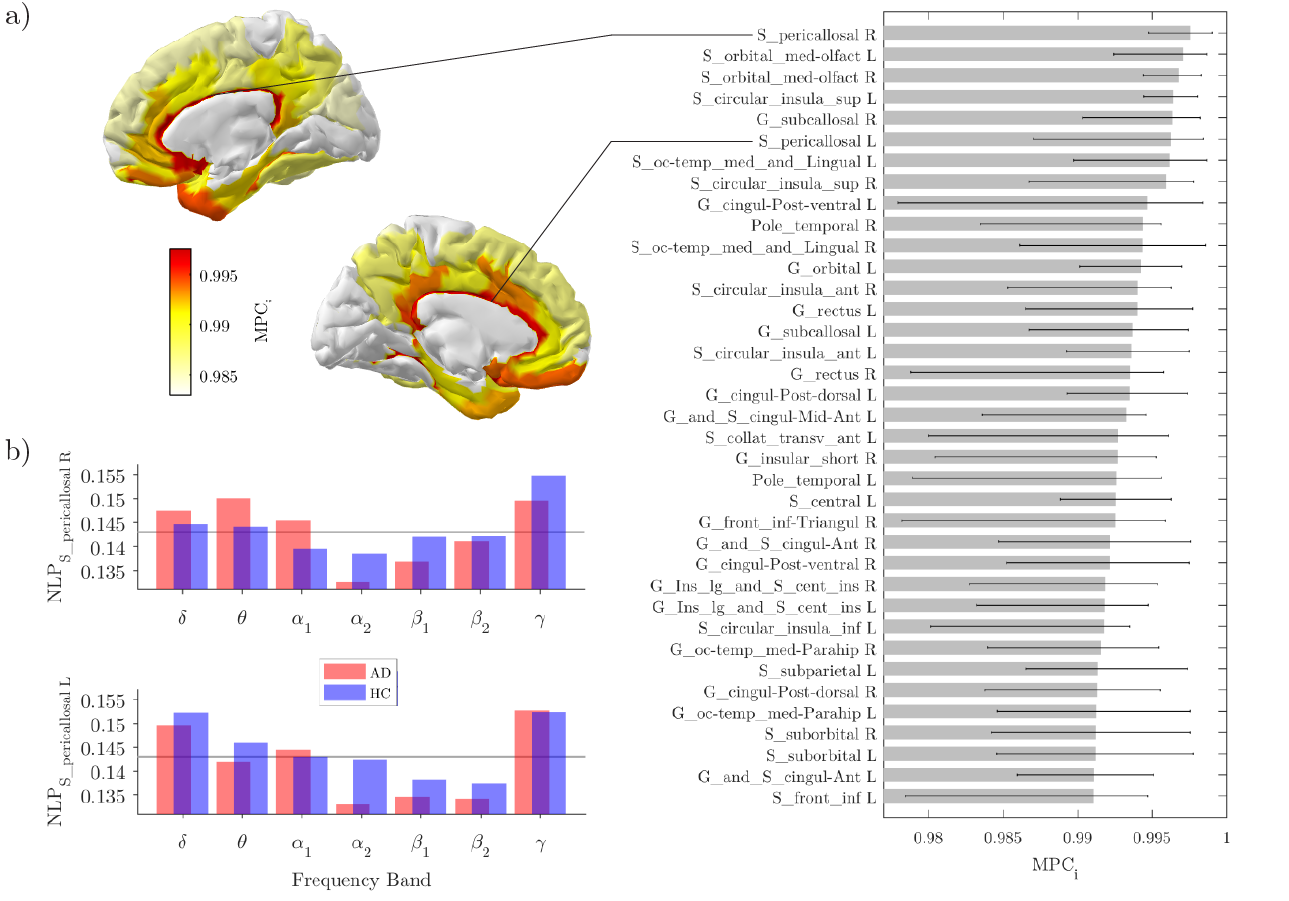}
	\caption{Inter-frequency hub centrality distribution. a) The median values of local multi-participation coefficients ($MPC_i$) are shown over the cortical surface for the healthy group. Only the top $25\%$ is illustrated for the sake of visualization. The corresponding list of ROIs is illustrated in the horizontal bar plot. b) Group-median values of the node-degree layer proportion ($NLP_i$) for the right and left cingulate cortex. The grey line corresponds to the expected value if connectivity were equally distributed across frequency bands ($NLP_i=1/7$).}
	\label{fig:mpc}
\end{figure}

\newpage
\begin{figure}[!ht]
	\centering
	\includegraphics[width=14cm]{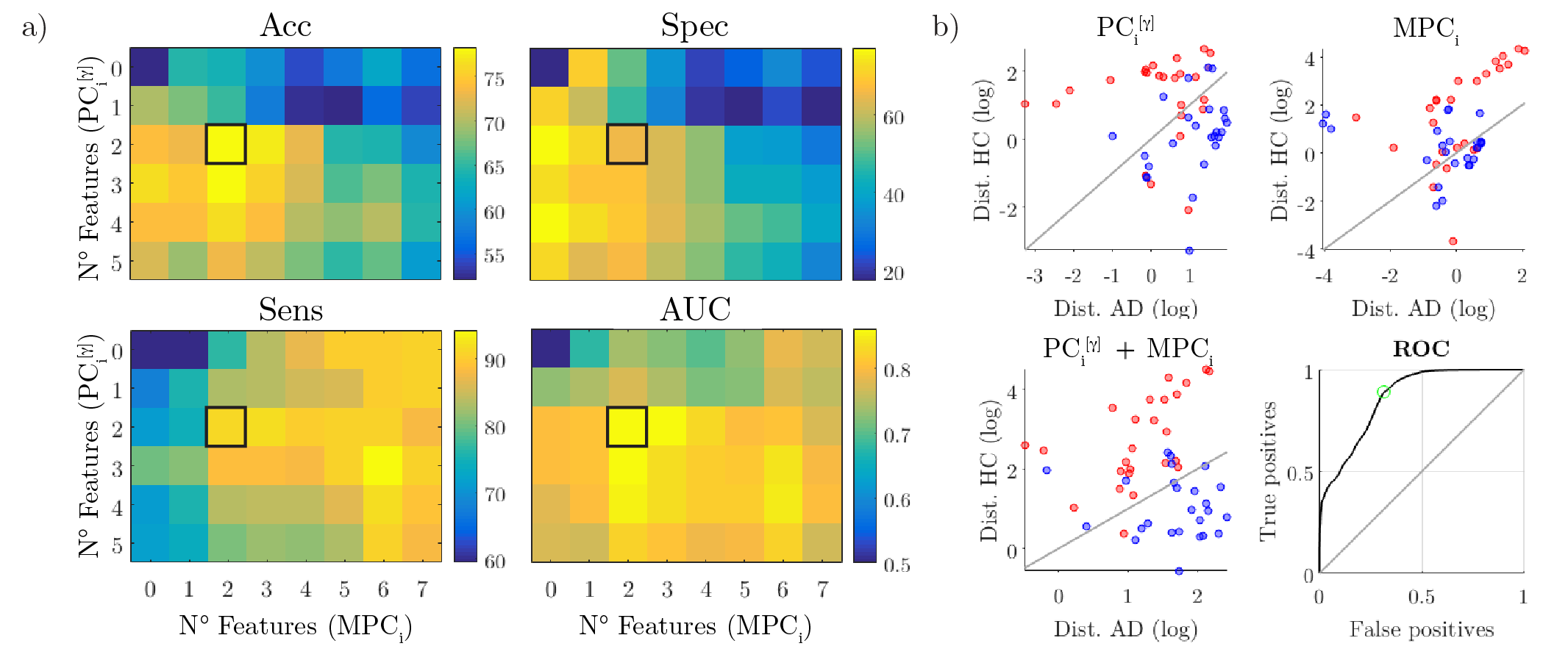}
	\caption{Classification performance of brain network features.
	a) Matrices show the classification rates (accuracy=Acc, specificity=Spec, sensitivity=Sens, area under the curve=AUC) corresponding to the combination of the most significant $PC^{[\gamma]}_i$ and $MPC_i$ network features, respectively on the rows and columns of each matrix. Black squares highlight the highest accuracy rate and the corresponding specificity, sensitivity and AUC.
	b) Scatter plots show the Mahalanobis distance of each subject from the $AD$ and $HC$ classes.  Separation lines ($y=x$: equal distances) are drawn in grey. Red circles stand for Alzheimer's disease (AD) subjects , blue ones for healthy controls (HC).
The bottom right plot shows the ROC curve associated with the best network features configuration. The optimal point is marked by a green circle.}
	\label{fig:classification}
\end{figure}

\newpage
\begin{figure}[!ht]
	\centering
	\includegraphics[width=14cm]{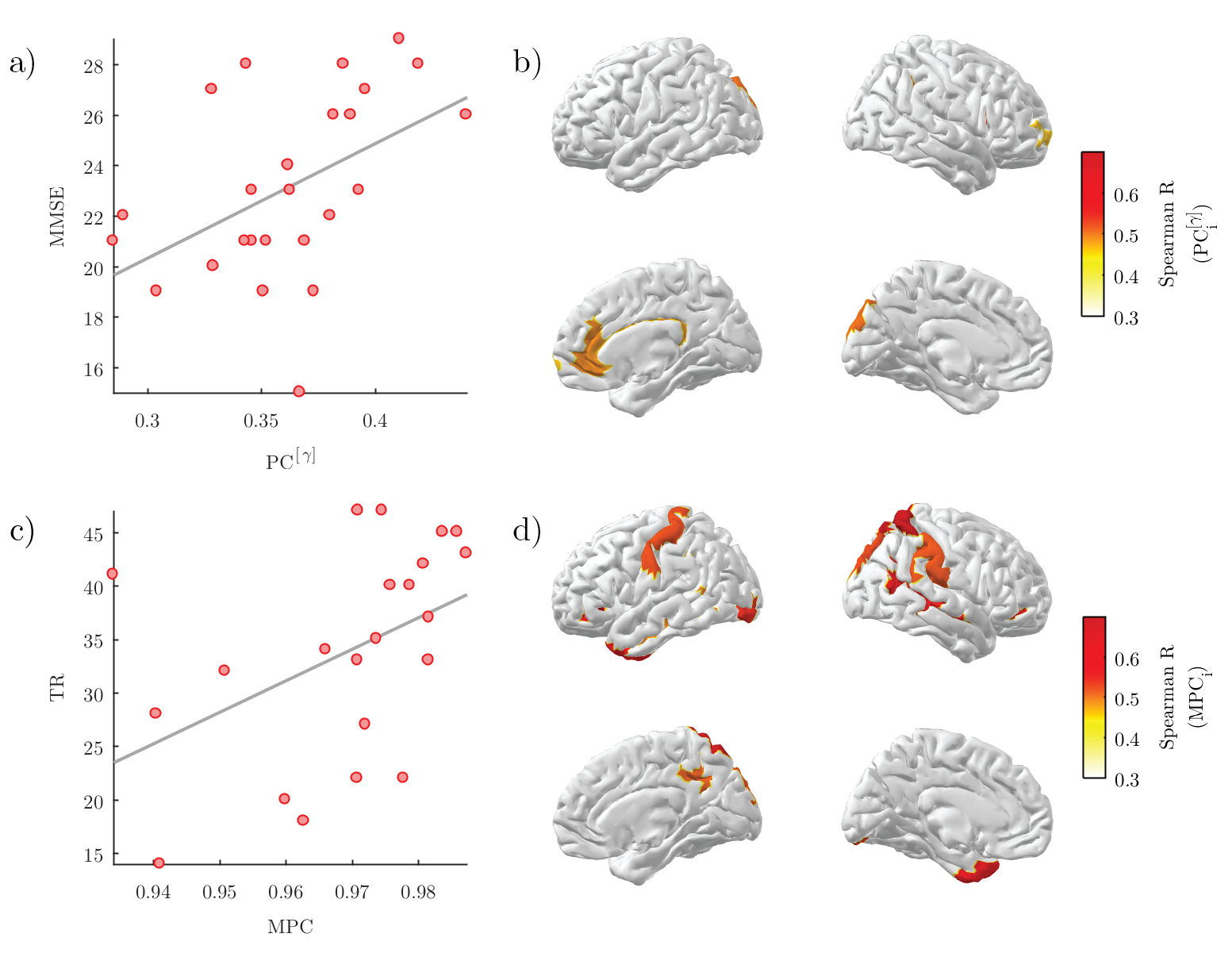}
	\caption{Correlation between brain network properties and cognitive/memory scores.
	a) Scatter plot of the global participation coefficient in the \textit{gamma} band ($PC^{[\gamma]}$) and the mini-mental state examination (MMSE) score of AD subjects (Spearman's correlation $R = 0.4909$, $p = 0.0127$).
	b) Correlation brain maps of the local participation coefficient in the \textit{gamma} band ($PC_i^{[\gamma]}$) and the mini-mental state examination (MMSE) score of AD subjects. Only significant $R$ values are illustrated ($p<0.05$, FDR corrected).
	c) Scatter plot of the global multi-participation coefficient ($PC$) and the total recall (TR) score of AD subjects (Spearman's correlation  $R = 0.5547$, $p = 0.0074$).
	d) Correlation brain maps of the local multi-participation coefficient ($MPC_i$) and the total recall (TR) score of AD subjects. Only significant $R$ values are illustrated ($p<0.05$, FDR corrected).
}
	\label{fig:correlations}
\end{figure}

%%%%%%%%%%%%%%%%%%%%%%%%%%%%%%%%%%%%%%%%%%%%%%%%%%%%%%%%%%%%

\newpage
\begin{table}[!ht]
\footnotesize
	\centering
	\begin{tabular}{lllll}
		\hline
		     & Control (HC) & Alzheimer (AD) & $p$-value  \\ \hline
		% Gender (M/F) & 7/18       & 12/13       & ${0.01}$     \\
		Age  & 70.8 (9.1)   & 73.5 (9.4)     & $0.3142$   \\
		MMSE & 28.2 (1.4)   & 23.2 (3.6)     & $<10^{-5}$ \\
		FR   & 31.5 (6.6)   & 14.9 (6.5)     & $<10^{-5}$ \\
		% CR   & 14.9 (6.4) & 19.0 (7.5)  & $0.0481$   \\
		TR   & 46.3 (1.5)   & 33.9 (10.0)    & $<10^{-5}$ \\
		% React        & 89.6 (8.6) & 61.8 (23.2) & $<10^{-5}$ \\ \hline
	\end{tabular}
	\caption{Characteristics, cognitive and memory scores of experimental subjects. Mean values and standard deviations (between parentheses) are reported. The last column shows the $p$-values returned by a non-parametric permutation t-tests with $10\,000$ realizations. MMSE = mini-mental state examination score; TR = total recall memory test score (/48); FR = free recall memory test (/48).}
	\label{tab:clinical_data}
\end{table}

\newpage
\begin{table}[!ht]
\footnotesize

	\centering
	\begin{tabular}{rrllrr}
		\hline
		Index                              & Rank & ROI label                     & Cortex        & $Z$ score & $p$-value \\ \hline
		\multirow{5}{*}{$PC_i^{[\gamma]}$} & 1    & Lat\_Fis-ant-Horizont L       & Frontal       & -3.6507   & 0.0007    \\
		                                   & 2    & Pole\_temporal R              & Temporal      & -2.8642   & 0.0063    \\
		                                   & 3    & G\_front\_inf-Triangul L      & Frontal       & -2.4562   & 0.0198    \\
		                                   & 4    & \textbf{S\_temporal\_transverse L}     & Temporal      & -2.3887   & 0.0207    \\
		                                   & 5    & \textbf{G\_pariet\_inf-Supramar L}     & Parietal      & -2.3820   & 0.0222    \\ \hline
		\multirow{7}{*}{$MPC_i$}           & 1    & G\_precentral R               & Motor & -3.4735   & 0.0006    \\
		                                   & 2    & G\_front\_inf-Opercular R     &  Motor & -2.5239   & 0.0127    \\
		                                   & 3    & S\_oc\_middle\_and\_Lunatus L & Occipital     & -2.4582   & 0.0138    \\
		                                   & 4    & \textbf{G\_pariet\_inf-Supramar L}     & Parietal      & -2.4860   & 0.0142    \\
		                                   & 5    & \textbf{S\_interm\_prim-Jensen L}      & Parietal      & -2.3708   & 0.0147    \\
		                                   & 6    & \textbf{S\_temporal\_transverse R}     & Temporal      & -2.3996   & 0.0191    \\
		                                   & 7    & \textbf{S\_pericallosal R }            & Limbic        & -2.3041   & 0.0203    \\ \hline
	\end{tabular}
	\caption{Statistical group differences for local brain network properties. ROI labels, abbreviated according to the Destrieux atlas, are ranked according to the resulting $p$-values. The same ranks are used as labels in \autoref{fig:participation}. ROIs highlighted in bold belong to the default mode network (DMN).}
	\label{tab:local_participation}
\end{table}

\newpage
\begin{table}[!ht]
\footnotesize
	\centering
	\begin{tabular}{rrllrr}
		\hline
		Correlation                                 & Rank & ROI label                               & Cortex          & $R$ coeff. & $p$-value \\ \hline
		\multirow{6}{*}{$PC_i^{[\gamma]}$ - MMSE} & 1    & Lat\_Fis-ant-Vertical R                 & Frontal         & 0.5480          & 0.0046    \\
		                                            & 2    & G\_occipital\_sup L                     & Occipital       & 0.5005          & 0.0108    \\
		                                            & 3    & \textbf{S\_interm\_prim-Jensen R}                & Parietal        & 0.4948          & 0.0119    \\
		                                            & 4    & \textbf{G\_and\_S\_cingul-Ant R}        & Limbic         & 0.4864          & 0.0137    \\
		                                            & 5    & \textbf{S\_pericallosal R}                       & Limbic          & 0.4735          & 0.0168    \\
		                                            & 6    & G\_and\_S\_transv\_frontopol R & Frontal         & 0.4585          & 0.0212    \\ \hline

		\multirow{15}{*}{$MPC_i$ - TR}            & 1    & Lat\_Fis-ant-Horizont L        & Frontal         & 0.6915          & 0.0004    \\
		                                            & 2    & S\_collat\_transv\_post L               & Occipital       & 0.6706          & 0.0006    \\
		                                            & 3    & S\_circular\_insula\_ant L              & Frontal         & 0.6214          & 0.0020    \\
		                                            & 4    & \textbf{G\_parietal\_sup R}                      & Parietal        & 0.6061          & 0.0028    \\
		                                            & 5    & S\_orbital\_lateral R          & Frontal         & 0.5920          & 0.0037    \\
		                                            & 6    & Pole\_temporal L               & Temporal        & 0.5739          & 0.0052    \\
		                                            & 7    & S\_orbital\_lateral L          & Frontal         & 0.5462          & 0.0085    \\
		                                            & 8    & \textbf{S\_temporal\_sup R}             & Temporal        & 0.5457          & 0.0086    \\
		                                            & 9    & G\_and\_S\_occipital\_inf L             & Occipital       & 0.5368          & 0.0100    \\
		                                            & 10   & G\_occipital\_sup R                     & Occipital       & 0.5208          & 0.0130    \\
		                                            & 11   & G\_postcentral L                        & Sensory & 0.5191          & 0.0133    \\
		                                            & 12   & \textbf{G\_pariet\_inf-Supramar R}      & Parietal        & 0.5151          & 0.0142    \\
		                                            & 13   & \textbf{S\_subparietal R}               & Parietal        & 0.5066          & 0.0161    \\
		                                            & 14   & \textbf{S\_interm\_prim-Jensen L}                & Parietal        & 0.4915          & 0.0202    \\
		                                            & 15   & \textbf{S\_temporal\_inf L}             & Temporal        & 0.4869          & 0.0216    \\ \hline
	\end{tabular}
	\caption{Correlations of local brain network properties and cognitive/memory scores. ROI labels, abbreviated according to the Destrieux atlas, are ranked according to the resulting $p$-values. ROIs written in bold belong to the default mode network (DMN).}
	\label{tab:local_correlation}
\end{table}

%% file: TextS.tex
% !TEX root = Guillon2017_arxiv.tex

%\documentclass[1p]{elsarticle}
%
%\usepackage{natbib}
%\usepackage{lineno}
%\usepackage[breaklinks=true]{hyperref}
%\usepackage{xr}
%\usepackage{amsmath}
%\usepackage{amssymb}
%\usepackage{booktabs}
%\usepackage{mathtools}
%\usepackage{multirow}
%\usepackage{graphicx}
%\usepackage{color}
%\usepackage{soul}
%\usepackage{setspace}
%\usepackage[author={Jeremy Guillon}]{pdfcomment}
%
%\setcitestyle{aysep={}}
%\newcommand{\bigcdot}{\raisebox{-0.25ex}{\scalebox{1.2}{$\cdot$}}}
%\newcommand{\Hz}{\ensuremath\text{Hz}}
%\definecolor{yellow}{rgb}{1,1,0}
%\newcommand{\note}{\pdfcomment[color=yellow]}
%\renewcommand*{\figureautorefname}{Fig.}
%\renewcommand*{\tableautorefname}{Tab.}
%\renewcommand*{\equationautorefname}{Eq.}
%%\linespread{1}
%
%\DeclarePairedDelimiter\abs{\lvert}{\rvert}
%\DeclarePairedDelimiter\norm{\lVert}{\rVert}
%
%
%\modulolinenumbers[1]
%\bibliographystyle{elsarticle-num} % Harvard Bib. Style
%%\bibliographystyle{elsarticle-num} % Harvard Bib. Style
%\biboptions{round,sort&compress}
%%\bibliographystyle{CSE} % Harvard Bib. Style
%%\biboptions{authoryear,round,sort&compress}

\renewcommand{\thefigure}{S\arabic{figure}}
\renewcommand{\thetable}{S\arabic{table}}
\renewcommand{\theequation}{S\arabic{equation}}

\setcounter{figure}{0}
\setcounter{table}{0}
\setcounter{equation}{0}

%\begin{document}
\subsection*{Supplementary Text} \label{subsec:TextS}

%\subsubsection*{Coefficient of variation}
The global coefficient of variation is given by averaging $CV_i$ values across all the nodes:
\begin{equation}
	CV = \frac{1}{n} \sum_{i=1}^{N} CV_i
	= \frac{1}{n} \sum_{i=1}^{N} \frac{\sigma_{k_i}^{[\bigcdot]}}{k_i^{[\bigcdot]}}
	\label{eq:cv}
\end{equation}
where $\sigma_{k_i}^{[\bigcdot]}$ is the standard deviation of the degree of node $i$ across layers and $k_i^{[\bigcdot]}$ is the mean value.

Differently from $MPC$, $CV$ tends to $0$ when the links of the nodes tend to evenly distribute across layers, and give higher values when they rather tend to be concentrated in one layer or, more in general, differently distributed across layers.

%\end{document}

%% file: SupplementaryMaterial.tex
% !TEX root = Guillon2017_arxiv.tex

% \subsection{Correlation with the MMSE}

% There is no correletation between the MMSE and the \textit{global} multi-participation coefficient (Figure 5.a). Locally some ROIs show either a positive correlation (?, ?, ?) or a negative correlation (...). We can note that the positively correlated ROIs also have a significantly higher multi-participation coefficient in controls (Figure 4.b).
% The small blue region isn't significant in terms of difference of MPC, so its correlation might be "rally linked to the MMSE". Because a correlation associated to a significantly different ROI is easly explained since all the HCs have a hight MPC, and all the ADs a low one... If we plot the correlation curve, it should look like two stairs steps overlaped by a positive linear curve.

\renewcommand{\thefigure}{S\arabic{figure}}
\renewcommand{\thetable}{S\arabic{table}}
\renewcommand{\theequation}{S\arabic{equation}}

\setcounter{figure}{0}
\setcounter{table}{0}
\setcounter{equation}{0}

\subsection*{Supplementary Figures} \label{subsec:FigS}

\begin{figure}[!ht]
	\centering
	\includegraphics[width=10cm]{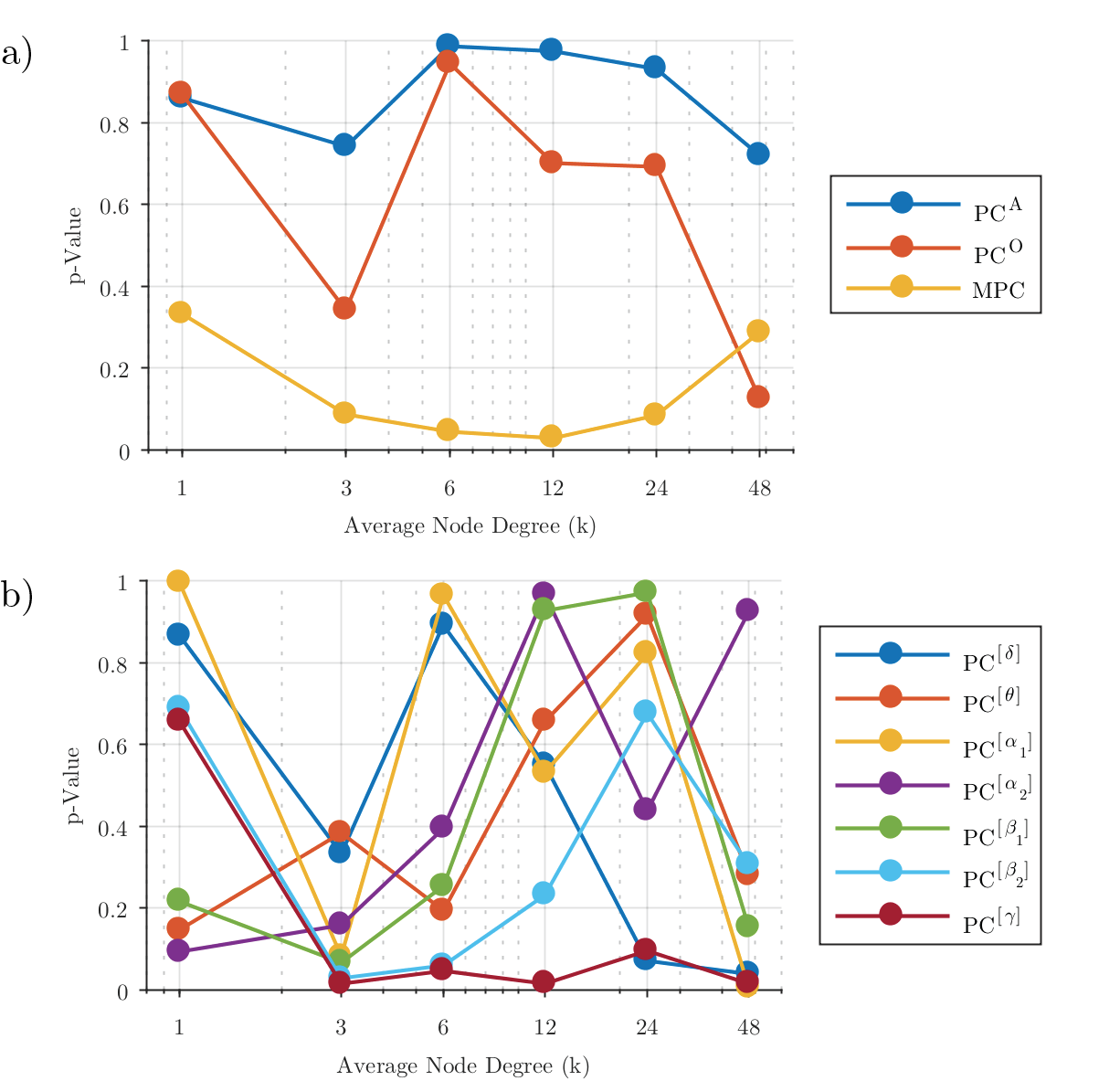}
	\caption{Statistical differences between global brain network properties of AD and HC subjects.
	These figures illustrate the $p$-values resulting from the permutation t-tests as a function of the average node degree $k$ used to threshold the layers of the multi-frequency brain networks. In panel a), we show the $p$-values for multi-layer and flattened analysis whereas in panel b) the $p$-values resulting from single-layer analysis.}
	\label{fig:thresholding}
\end{figure}

\newpage
\begin{figure}[!ht]
	\centering
	\includegraphics[width=10cm]{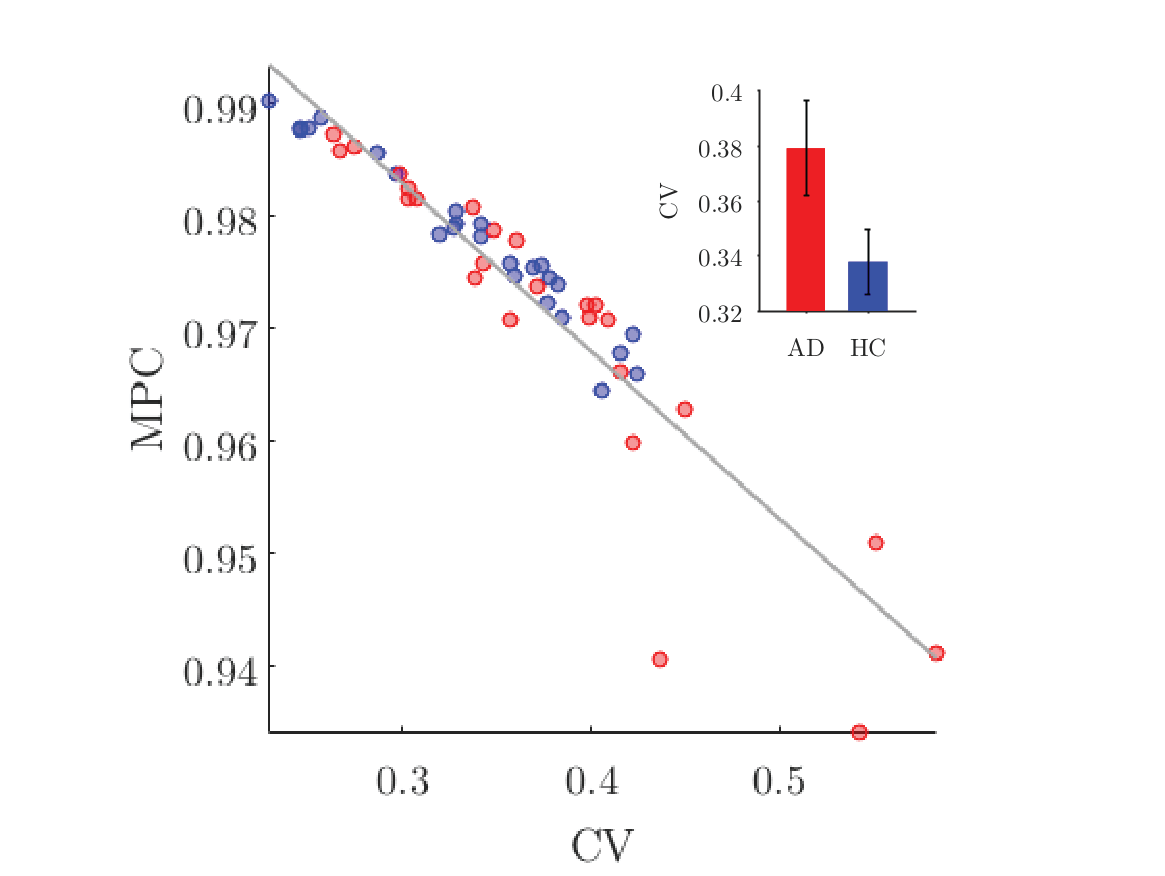}
	\caption{This figure shows the global coefficient of variation ($CV$): first the difference between the populations as an inset plot ($p=0.0521$) and the correlation with the global multi-participation coefficient ($MPC$) as a main plot ($p<10^{-15}$, $R=-0.9742$).}
	\label{fig:coefficient_of_variation}
\end{figure}

\newpage
\begin{figure}[!ht]
	\centering
	\includegraphics[width=14cm]{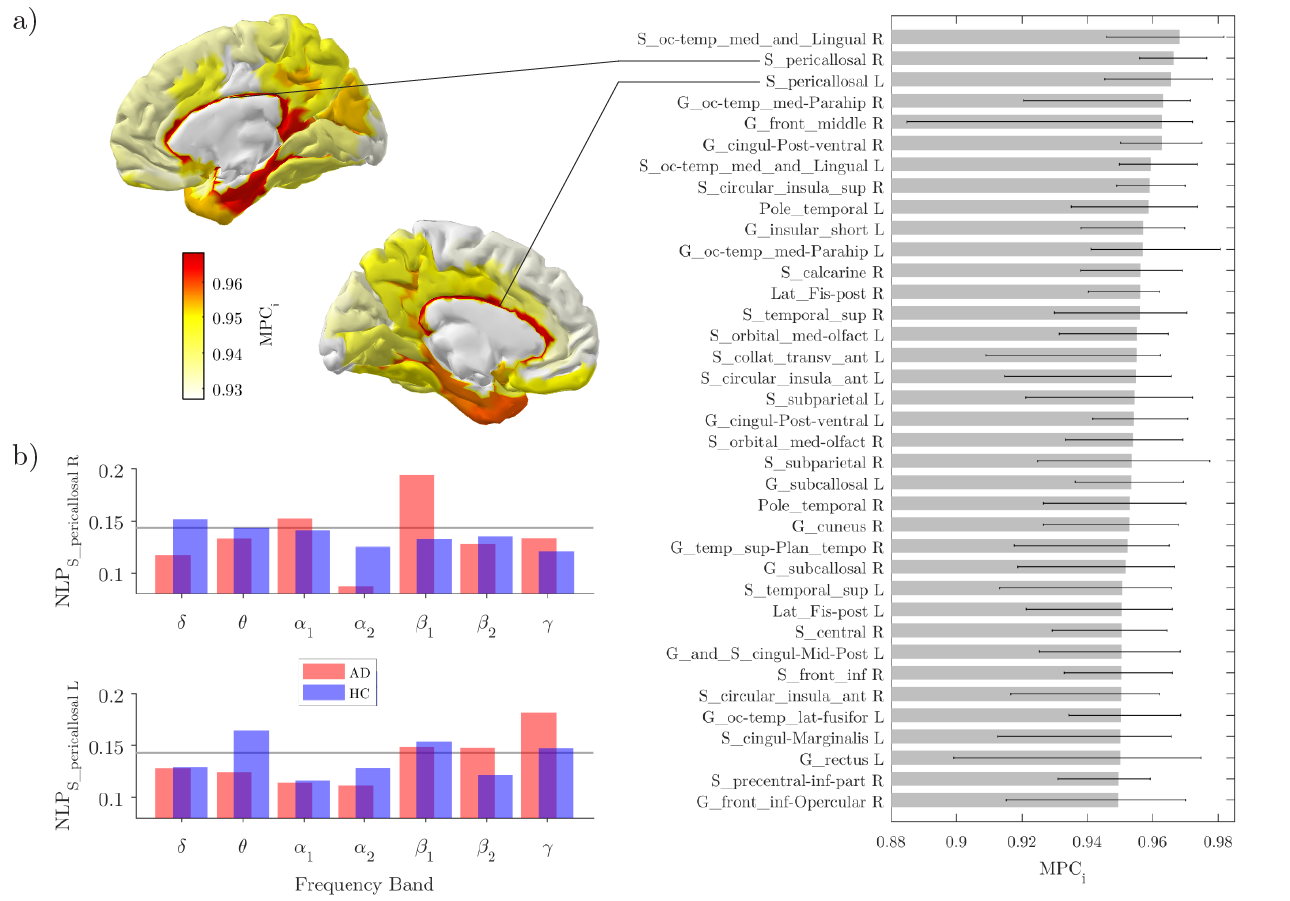}
	\caption{Inter-frequency hub centrality distribution for brain networks obtained with imaginary coherence. a) The median values of local multi-participation coefficients ($MPC_i$) are shown over the cortical surface for the healthy group. Only the top $25\%$ is illustrated for the sake of visualization. The corresponding list of ROIs is illustrated in the horizontal bar plot. b) Group-median values of the node-degree layer proportion ($NLP_i$) for the right and left cingulate cortex. The grey line corresponds to the expected value if connectivity were equally distributed across frequency bands ($NLP=1/7$).}
	\label{fig:mpc_imcoh}
\end{figure}

%% file: Guillon2017_arxiv.bbl
\begin{thebibliography}{10}
\expandafter\ifx\csname url\endcsname\relax
  \def\url#1{\texttt{#1}}\fi
\expandafter\ifx\csname urlprefix\endcsname\relax\def\urlprefix{URL }\fi
\expandafter\ifx\csname href\endcsname\relax
  \def\href#1#2{#2} \def\path#1{#1}\fi

\bibitem{stam_modern_2014}
C.~J. Stam,
  \href{http://www.nature.com/nrn/journal/v15/n10/abs/nrn3801.html}{Modern
  network science of neurological disorders}, Nat Rev Neurosci 15~(10) (2014)
  683--695.
\newblock \href {http://dx.doi.org/10.1038/nrn3801}
  {\path{doi:10.1038/nrn3801}}.

\bibitem{tijms_alzheimers_2013}
B.~M. Tijms, A.~M. Wink, W.~{de Haan}, W.~M. {van der Flier}, C.~J. Stam,
  P.~Scheltens, F.~Barkhof, Alzheimer's disease: Connecting findings from graph
  theoretical studies of brain networks, Neurobiol. Aging 34~(8) (2013)
  2023--2036.
\newblock \href {http://dx.doi.org/10.1016/j.neurobiolaging.2013.02.020}
  {\path{doi:10.1016/j.neurobiolaging.2013.02.020}}.

\bibitem{stam_use_2010}
C.~J. Stam,
  \href{http://www.sciencedirect.com/science/article/pii/S0022510X09007849}{Use
  of magnetoencephalography ({{MEG}}) to study functional brain networks in
  neurodegenerative disorders}, Journal of the Neurological Sciences
  289~(1\textendash{}2) (2010) 128--134.
\newblock \href {http://dx.doi.org/10.1016/j.jns.2009.08.028}
  {\path{doi:10.1016/j.jns.2009.08.028}}.

\bibitem{wenk_neuropathologic_2003}
G.~L. Wenk, Neuropathologic changes in {{Alzheimer}}'s disease, J Clin
  Psychiatry 64 Suppl 9 (2003) 7--10.

\bibitem{rose_loss_2000}
S.~E. Rose, F.~Chen, J.~B. Chalk, F.~O. Zelaya, W.~E. Strugnell, M.~Benson,
  J.~Semple, D.~M. Doddrell, Loss of connectivity in {{Alzheimer}}'s disease:
  An evaluation of white matter tract integrity with colour coded {{MR}}
  diffusion tensor imaging, J. Neurol. Neurosurg. Psychiatr. 69~(4) (2000)
  528--530.

\bibitem{zhou_abnormal_2008}
Y.~Zhou, J.~H. Dougherty, K.~F. Hubner, B.~Bai, R.~L. Cannon, R.~K. Hutson,
  Abnormal connectivity in the posterior cingulate and hippocampus in early
  {{Alzheimer}}'s disease and mild cognitive impairment, Alzheimers Dement
  4~(4) (2008) 265--270.
\newblock \href {http://dx.doi.org/10.1016/j.jalz.2008.04.006}
  {\path{doi:10.1016/j.jalz.2008.04.006}}.

\bibitem{lo_diffusion_2010}
C.-Y. Lo, P.-N. Wang, K.-H. Chou, J.~Wang, Y.~He, C.-P. Lin, Diffusion tensor
  tractography reveals abnormal topological organization in structural cortical
  networks in {{Alzheimer}}'s disease, J. Neurosci. 30~(50) (2010)
  16876--16885.
\newblock \href {http://dx.doi.org/10.1523/JNEUROSCI.4136-10.2010}
  {\path{doi:10.1523/JNEUROSCI.4136-10.2010}}.

\bibitem{sanz-arigita_loss_2010-1}
E.~J. Sanz-Arigita, M.~M. Schoonheim, J.~S. Damoiseaux, S.~A. R.~B. Rombouts,
  E.~Maris, F.~Barkhof, P.~Scheltens, C.~J. Stam,
  \href{http://journals.plos.org/plosone/article?id=10.1371/journal.pone.0013788}{Loss
  of `{{Small}}-{{World}}' {{Networks}} in {{Alzheimer}}'s {{Disease}}: {{Graph
  Analysis}} of {{fMRI Resting}}-{{State Functional Connectivity}}}, PLOS ONE
  5~(11) (2010) e13788.
\newblock \href {http://dx.doi.org/10.1371/journal.pone.0013788}
  {\path{doi:10.1371/journal.pone.0013788}}.

\bibitem{stam_graph_2009}
C.~J. Stam, W.~de~Haan, A.~Daffertshofer, B.~F. Jones, I.~Manshanden, A.~M.
  v.~C. van Walsum, T.~Montez, J.~P.~A. Verbunt, J.~C. de~Munck, B.~W. van
  Dijk, H.~W. Berendse, P.~Scheltens,
  \href{http://brain.oxfordjournals.org/content/132/1/213}{Graph theoretical
  analysis of magnetoencephalographic functional connectivity in
  {{Alzheimer}}'s disease}, Brain 132~(1) (2009) 213--224.
\newblock \href {http://dx.doi.org/10.1093/brain/awn262}
  {\path{doi:10.1093/brain/awn262}}.

\bibitem{de_haan_functional_2009}
W.~{de Haan}, Y.~A. Pijnenburg, R.~L. Strijers, Y.~{van der Made}, W.~M. {van
  der Flier}, P.~Scheltens, C.~J. Stam,
  \href{http://dx.doi.org/10.1186/1471-2202-10-101}{Functional neural network
  analysis in frontotemporal dementia and {{Alzheimer}}'s disease using {{EEG}}
  and graph theory}, BMC Neuroscience 10 (2009) 101.
\newblock \href {http://dx.doi.org/10.1186/1471-2202-10-101}
  {\path{doi:10.1186/1471-2202-10-101}}.

\bibitem{miraglia_searching_2017}
F.~Miraglia, F.~Vecchio, P.~M. Rossini,
  \href{http://www.sciencedirect.com/science/article/pii/S0166432816307136}{Searching
  for signs of aging and dementia in {{EEG}} through network analysis},
  Behavioural Brain Research 317 (2017) 292--300.
\newblock \href {http://dx.doi.org/10.1016/j.bbr.2016.09.057}
  {\path{doi:10.1016/j.bbr.2016.09.057}}.

\bibitem{bassett_dynamic_2011}
D.~S. Bassett, N.~F. Wymbs, M.~A. Porter, P.~J. Mucha, J.~M. Carlson, S.~T.
  Grafton, \href{http://www.pnas.org/content/108/18/7641}{Dynamic
  reconfiguration of human brain networks during learning}, PNAS 108~(18)
  (2011) 7641--7646.
\newblock \href {http://dx.doi.org/10.1073/pnas.1018985108}
  {\path{doi:10.1073/pnas.1018985108}}.

\bibitem{crossley_hubs_2014}
N.~A. Crossley, A.~Mechelli, J.~Scott, F.~Carletti, P.~T. Fox, P.~McGuire,
  E.~T. Bullmore,
  \href{http://www.ncbi.nlm.nih.gov/pmc/articles/PMC4107735/}{The hubs of the
  human connectome are generally implicated in the anatomy of brain disorders},
  Brain 137~(8) (2014) 2382--2395.
\newblock \href {http://dx.doi.org/10.1093/brain/awu132}
  {\path{doi:10.1093/brain/awu132}}.

\bibitem{buckner_cortical_2009}
R.~L. Buckner, J.~Sepulcre, T.~Talukdar, F.~M. Krienen, H.~Liu, T.~Hedden,
  J.~R. Andrews-Hanna, R.~A. Sperling, K.~A. Johnson, Cortical hubs revealed by
  intrinsic functional connectivity: Mapping, assessment of stability, and
  relation to {{Alzheimer}}'s disease, J. Neurosci. 29~(6) (2009) 1860--1873.
\newblock \href {http://dx.doi.org/10.1523/JNEUROSCI.5062-08.2009}
  {\path{doi:10.1523/JNEUROSCI.5062-08.2009}}.

\bibitem{de_haan_disrupted_2012}
W.~{de Haan}, W.~M. {van der Flier}, T.~Koene, L.~L. Smits, P.~Scheltens, C.~J.
  Stam,
  \href{http://www.sciencedirect.com/science/article/pii/S1053811911013371}{Disrupted
  modular brain dynamics reflect cognitive dysfunction in {{Alzheimer}}'s
  disease}, NeuroImage 59~(4) (2012) 3085--3093.
\newblock \href {http://dx.doi.org/10.1016/j.neuroimage.2011.11.055}
  {\path{doi:10.1016/j.neuroimage.2011.11.055}}.

\bibitem{engels_declining_2015}
M.~M. Engels, C.~J. Stam, W.~M. {van der Flier}, P.~Scheltens, H.~{de Waal},
  E.~C. {van Straaten},
  \href{http://www.ncbi.nlm.nih.gov/pmc/articles/PMC4545875/}{Declining
  functional connectivity and changing hub locations in {{Alzheimer}}'s
  disease: An {{EEG}} study}, BMC Neurol 15.
\newblock \href {http://dx.doi.org/10.1186/s12883-015-0400-7}
  {\path{doi:10.1186/s12883-015-0400-7}}.

\bibitem{de_vico_fallani_graph_2014}
F.~De~Vico~Fallani, J.~Richiardi, M.~Chavez, S.~Achard,
  \href{http://rstb.royalsocietypublishing.org/content/369/1653/20130521}{Graph
  analysis of functional brain networks: Practical issues in translational
  neuroscience}, Phil. Trans. R. Soc. B 369~(1653) (2014) 20130521.
\newblock \href {http://dx.doi.org/10.1098/rstb.2013.0521}
  {\path{doi:10.1098/rstb.2013.0521}}.

\bibitem{bullmore_complex_2009}
E.~Bullmore, O.~Sporns, Complex brain networks: Graph theoretical analysis of
  structural and functional systems, Nat. Rev. Neurosci. 10~(3) (2009)
  186--198.
\newblock \href {http://dx.doi.org/10.1038/nrn2575}
  {\path{doi:10.1038/nrn2575}}.

\bibitem{canolty_functional_2010}
R.~T. Canolty, R.~T. Knight,
  \href{http://www.sciencedirect.com/science/article/pii/S1364661310002068}{The
  functional role of cross-frequency coupling}, Trends in Cognitive Sciences
  14~(11) (2010) 506--515.
\newblock \href {http://dx.doi.org/10.1016/j.tics.2010.09.001}
  {\path{doi:10.1016/j.tics.2010.09.001}}.

\bibitem{jirsa_cross-frequency_2013}
V.~Jirsa, V.~M{\"u}ller,
  \href{http://www.ncbi.nlm.nih.gov/pmc/articles/PMC3699761/}{Cross-frequency
  coupling in real and virtual brain networks}, Front Comput Neurosci 7.
\newblock \href {http://dx.doi.org/10.3389/fncom.2013.00078}
  {\path{doi:10.3389/fncom.2013.00078}}.

\bibitem{brookes_multi-layer_2016-1}
M.~J. Brookes, P.~K. Tewarie, B.~A.~E. Hunt, S.~E. Robson, L.~E. Gascoyne,
  E.~B. Liddle, P.~F. Liddle, P.~G. Morris,
  \href{http://www.sciencedirect.com/science/article/pii/S1053811916001543}{A
  multi-layer network approach to {{MEG}} connectivity analysis}, NeuroImage
  132 (2016) 425--438.
\newblock \href {http://dx.doi.org/10.1016/j.neuroimage.2016.02.045}
  {\path{doi:10.1016/j.neuroimage.2016.02.045}}.

\bibitem{fraga_characterizing_2013}
F.~J. Fraga, T.~H. Falk, P.~A.~M. Kanda, R.~Anghinah,
  \href{http://www.ncbi.nlm.nih.gov/pmc/articles/PMC3754998/}{Characterizing
  {{Alzheimer}}'s {{Disease Severity}} via {{Resting}}-{{Awake EEG Amplitude
  Modulation Analysis}}}, PLoS One 8~(8).
\newblock \href {http://dx.doi.org/10.1371/journal.pone.0072240}
  {\path{doi:10.1371/journal.pone.0072240}}.

\bibitem{blinowska_functional_2016}
K.~J. Blinowska, F.~Rakowski, M.~Kaminski, F.~De~Vico~Fallani, C.~Del~Percio,
  R.~Lizio, C.~Babiloni, Functional and effective brain connectivity for
  discrimination between {{Alzheimer}}'s patients and healthy individuals:
  {{A}} study on resting state {{EEG}} rhythms, Clin Neurophysiol\href
  {http://dx.doi.org/10.1016/j.clinph.2016.10.002}
  {\path{doi:10.1016/j.clinph.2016.10.002}}.

\bibitem{ghanbari_functionally_2014}
Y.~Ghanbari, L.~Bloy, V.~Shankar, J.~C. Edgar, T.~P.~L. Roberts, R.~T. Schultz,
  R.~Verma, Functionally driven brain networks using multi-layer graph
  clustering, Med Image Comput Comput Assist Interv 17~(Pt 3) (2014) 113--120.

\bibitem{simas_algebraic_2015}
T.~Simas, M.~Chavez, P.~R. Rodriguez, A.~Diaz-Guilera,
  \href{http://www.ncbi.nlm.nih.gov/pmc/articles/PMC4491601/}{An algebraic
  topological method for multimodal brain networks comparisons}, Front Psychol
  6.
\newblock \href {http://dx.doi.org/10.3389/fpsyg.2015.00904}
  {\path{doi:10.3389/fpsyg.2015.00904}}.

\bibitem{battiston_multilayer_2016}
F.~Battiston, V.~Nicosia, M.~Chavez, V.~Latora,
  \href{http://arxiv.org/abs/1606.09115}{Multilayer motif analysis of brain
  networks}, arXiv:1606.09115 [cond-mat, physics:physics, q-bio]\href
  {http://arxiv.org/abs/1606.09115} {\path{arXiv:1606.09115}}.

\bibitem{de_domenico_mapping_2016}
M.~De~Domenico, S.~Sasai, A.~Arenas,
  \href{http://arxiv.org/abs/1603.05897}{Mapping multiplex hubs in human
  functional brain network}, arXiv:1603.05897 [cond-mat, physics:physics,
  q-bio]\href {http://arxiv.org/abs/1603.05897} {\path{arXiv:1603.05897}}.

\bibitem{battiston_structural_2014}
F.~Battiston, V.~Nicosia, V.~Latora,
  \href{http://link.aps.org/doi/10.1103/PhysRevE.89.032804}{Structural measures
  for multiplex networks}, Phys. Rev. E 89~(3) (2014) 032804.
\newblock \href {http://dx.doi.org/10.1103/PhysRevE.89.032804}
  {\path{doi:10.1103/PhysRevE.89.032804}}.

\bibitem{de_domenico_mathematical_2013}
M.~De~Domenico, A.~Sol{\'e}-Ribalta, E.~Cozzo, M.~Kivel{\"a}, Y.~Moreno, M.~A.
  Porter, S.~G{\'o}mez, A.~Arenas,
  \href{http://link.aps.org/doi/10.1103/PhysRevX.3.041022}{Mathematical
  {{Formulation}} of {{Multilayer Networks}}}, Phys. Rev. X 3~(4) (2013)
  041022.
\newblock \href {http://dx.doi.org/10.1103/PhysRevX.3.041022}
  {\path{doi:10.1103/PhysRevX.3.041022}}.

\bibitem{folstein_mini-mental_1975}
M.~F. Folstein, S.~E. Folstein, P.~R. McHugh, "{{Mini}}-mental state". {{A}}
  practical method for grading the cognitive state of patients for the
  clinician, J Psychiatr Res 12~(3) (1975) 189--198.

\bibitem{buschke_cued_1984}
H.~Buschke, \href{http://dx.doi.org/10.1080/01688638408401233}{Cued recall in
  {{Amnesia}}}, Journal of Clinical Neuropsychology 6~(4) (1984) 433--440.
\newblock \href {http://dx.doi.org/10.1080/01688638408401233}
  {\path{doi:10.1080/01688638408401233}}.

\bibitem{grober_screening_1988}
E.~Grober, H.~Buschke, H.~Crystal, S.~Bang, R.~Dresner, Screening for dementia
  by memory testing, Neurology 38~(6) (1988) 900--903.

\bibitem{pillon_explicit_1993}
B.~Pillon, B.~Deweer, Y.~Agid, B.~Dubois, Explicit memory in {{Alzheimer}}'s,
  {{Huntington}}'s, and {{Parkinson}}'s diseases, Arch. Neurol. 50~(4) (1993)
  374--379.

\bibitem{sarazin_amnestic_2007}
M.~Sarazin, C.~Berr, J.~De~Rotrou, C.~Fabrigoule, F.~Pasquier, S.~Legrain,
  B.~Michel, M.~Puel, M.~Volteau, J.~Touchon, M.~Verny, B.~Dubois, Amnestic
  syndrome of the medial temporal type identifies prodromal {{AD}}: A
  longitudinal study, Neurology 69~(19) (2007) 1859--1867.
\newblock \href {http://dx.doi.org/10.1212/01.wnl.0000279336.36610.f7}
  {\path{doi:10.1212/01.wnl.0000279336.36610.f7}}.

\bibitem{taulu_spatiotemporal_2006}
S.~Taulu, J.~Simola,
  \href{http://stacks.iop.org/0031-9155/51/i=7/a=008}{Spatiotemporal signal
  space separation method for rejecting nearby interference in {{MEG}}
  measurements}, Phys. Med. Biol. 51~(7) (2006) 1759.
\newblock \href {http://dx.doi.org/10.1088/0031-9155/51/7/008}
  {\path{doi:10.1088/0031-9155/51/7/008}}.

\bibitem{he_brain_1999}
B.~He, Brain electric source imaging: Scalp {{Laplacian}} mapping and cortical
  imaging, Crit Rev Biomed Eng 27~(3-5) (1999) 149--188.

\bibitem{baillet_evaluation_2001}
S.~Baillet, J.~J. Riera, G.~Marin, J.~F. Mangin, J.~Aubert, L.~Garnero,
  Evaluation of inverse methods and head models for {{EEG}} source localization
  using a human skull phantom, Phys Med Biol 46~(1) (2001) 77--96.

\bibitem{fischl_whole_2002}
B.~Fischl, D.~H. Salat, E.~Busa, M.~Albert, M.~Dieterich, C.~Haselgrove,
  A.~{van der Kouwe}, R.~Killiany, D.~Kennedy, S.~Klaveness, A.~Montillo,
  N.~Makris, B.~Rosen, A.~M. Dale, Whole brain segmentation: Automated labeling
  of neuroanatomical structures in the human brain, Neuron 33~(3) (2002)
  341--355.

\bibitem{fischl_sequence-independent_2004}
B.~Fischl, D.~H. Salat, A.~J.~W. {van der Kouwe}, N.~Makris, F.~S{\'e}gonne,
  B.~T. Quinn, A.~M. Dale, Sequence-independent segmentation of magnetic
  resonance images, Neuroimage 23 Suppl 1 (2004) S69--84.
\newblock \href {http://dx.doi.org/10.1016/j.neuroimage.2004.07.016}
  {\path{doi:10.1016/j.neuroimage.2004.07.016}}.

\bibitem{tadel_brainstorm:_2011}
F.~Tadel, S.~Baillet, J.~C. Mosher, D.~Pantazis, R.~M. Leahy, F.~Tadel,
  S.~Baillet, J.~C. Mosher, D.~Pantazis, R.~M. Leahy,
  \href{http://www.hindawi.com/journals/cin/2011/879716/abs/,\%0020http://www.hindawi.com/journals/cin/2011/879716/abs/}{Brainstorm:
  {{A User}}-{{Friendly Application}} for {{MEG}}/{{EEG Analysis}},
  {{Brainstorm}}: {{A User}}-{{Friendly Application}} for {{MEG}}/{{EEG
  Analysis}}}, Computational Intelligence and Neuroscience, Computational
  Intelligence and Neuroscience 2011, 2011 (2011) e879716.
\newblock \href
  {http://dx.doi.org/10.1155/2011/879716,\%002010.1155/2011/879716}
  {\path{doi:10.1155/2011/879716,\%002010.1155/2011/879716}}.

\bibitem{lin_assessing_2006}
F.-H. Lin, T.~Witzel, S.~P. Ahlfors, S.~M. Stufflebeam, J.~W. Belliveau, M.~S.
  H{\"a}m{\"a}l{\"a}inen,
  \href{http://www.sciencedirect.com/science/article/pii/S1053811905024973}{Assessing
  and improving the spatial accuracy in {{MEG}} source localization by
  depth-weighted minimum-norm estimates}, NeuroImage 31~(1) (2006) 160--171.
\newblock \href {http://dx.doi.org/10.1016/j.neuroimage.2005.11.054}
  {\path{doi:10.1016/j.neuroimage.2005.11.054}}.

\bibitem{destrieux_automatic_2010-1}
C.~Destrieux, B.~Fischl, A.~Dale, E.~Halgren,
  \href{http://www.ncbi.nlm.nih.gov/pmc/articles/PMC2937159/}{Automatic
  parcellation of human cortical gyri and sulci using standard anatomical
  nomenclature}, Neuroimage 53~(1) (2010) 1--15.
\newblock \href {http://dx.doi.org/10.1016/j.neuroimage.2010.06.010}
  {\path{doi:10.1016/j.neuroimage.2010.06.010}}.

\bibitem{carter_coherence_1987}
G.~C. Carter, Coherence and time delay estimation, Proceedings of the IEEE
  75~(2) (1987) 236--255.
\newblock \href {http://dx.doi.org/10.1109/PROC.1987.13723}
  {\path{doi:10.1109/PROC.1987.13723}}.

\bibitem{stam_generalized_2002}
C.~J. Stam, A.~M. {van Cappellen van Walsum}, Y.~A.~L. Pijnenburg, H.~W.
  Berendse, J.~C. {de Munck}, P.~Scheltens, B.~W. {van Dijk}, Generalized
  synchronization of {{MEG}} recordings in {{Alzheimer}}'s {{Disease}}:
  Evidence for involvement of the gamma band, J Clin Neurophysiol 19~(6) (2002)
  562--574.

\bibitem{babiloni_abnormal_2004}
C.~Babiloni, R.~Ferri, D.~V. Moretti, A.~Strambi, G.~Binetti, G.~Dal~Forno,
  F.~Ferreri, B.~Lanuzza, C.~Bonato, F.~Nobili, G.~Rodriguez, S.~Salinari,
  S.~Passero, R.~Rocchi, C.~J. Stam, P.~M. Rossini, Abnormal fronto-parietal
  coupling of brain rhythms in mild {{Alzheimer}}'s disease: A multicentric
  {{EEG}} study, Eur. J. Neurosci. 19~(9) (2004) 2583--2590.
\newblock \href {http://dx.doi.org/10.1111/j.0953-816X.2004.03333.x}
  {\path{doi:10.1111/j.0953-816X.2004.03333.x}}.

\bibitem{rubinov_complex_2010}
M.~Rubinov, O.~Sporns,
  \href{http://www.sciencedirect.com/science/article/pii/S105381190901074X}{Complex
  network measures of brain connectivity: {{Uses}} and interpretations},
  NeuroImage 52~(3) (2010) 1059--1069.
\newblock \href {http://dx.doi.org/10.1016/j.neuroimage.2009.10.003}
  {\path{doi:10.1016/j.neuroimage.2009.10.003}}.

\bibitem{guimera_cartography_2005}
R.~Guimer{\`a}, L.~A.~N. Amaral,
  \href{http://www.ncbi.nlm.nih.gov/pmc/articles/PMC2151742/}{Cartography of
  complex networks: Modules and universal roles}, J Stat Mech 2005~(P02001)
  (2005) P02001--1--P02001--13.
\newblock \href {http://dx.doi.org/10.1088/1742-5468/2005/02/P02001}
  {\path{doi:10.1088/1742-5468/2005/02/P02001}}.

\bibitem{newman_finding_2006}
M.~E.~J. Newman,
  \href{http://link.aps.org/doi/10.1103/PhysRevE.74.036104}{Finding community
  structure in networks using the eigenvectors of matrices}, Phys. Rev. E
  74~(3) (2006) 036104.
\newblock \href {http://dx.doi.org/10.1103/PhysRevE.74.036104}
  {\path{doi:10.1103/PhysRevE.74.036104}}.

\bibitem{de_vico_fallani_interhemispheric_2016}
F.~De~Vico~Fallani, S.~Clausi, M.~Leggio, M.~Chavez, M.~Valencia, A.~G.
  Maglione, F.~Babiloni, F.~Cincotti, D.~Mattia, M.~Molinari, Interhemispheric
  {{Connectivity Characterizes Cortical Reorganization}} in {{Motor}}-{{Related
  Networks After Cerebellar Lesions}}, Cerebellum\href
  {http://dx.doi.org/10.1007/s12311-016-0811-z}
  {\path{doi:10.1007/s12311-016-0811-z}}.

\bibitem{benjamini_controlling_1995}
Y.~Benjamini, Y.~Hochberg,
  \href{http://www.jstor.org/stable/2346101}{Controlling the {{False Discovery
  Rate}}: {{A Practical}} and {{Powerful Approach}} to {{Multiple Testing}}},
  Journal of the Royal Statistical Society. Series B (Methodological) 57~(1)
  (1995) 289--300.

\bibitem{zar_biostatistical_1999}
J.~H. Zar, Biostatistical Analysis, {Prentice Hall PTR}, 1999.

\bibitem{hastie_elements_2009}
T.~Hastie, R.~Tibshirani, J.~Friedman,
  \href{http://link.springer.com/10.1007/978-0-387-84858-7}{The {{Elements}} of
  {{Statistical Learning}}}, Springer Series in Statistics, {Springer New
  York}, New York, NY, 2009.

\bibitem{babiloni_mapping_2004}
C.~Babiloni, G.~Binetti, E.~Cassetta, D.~Cerboneschi, G.~Dal~Forno,
  C.~Del~Percio, F.~Ferreri, R.~Ferri, B.~Lanuzza, C.~Miniussi, D.~V. Moretti,
  F.~Nobili, R.~D. Pascual-Marqui, G.~Rodriguez, G.~L. Romani, S.~Salinari,
  F.~Tecchio, P.~Vitali, O.~Zanetti, F.~Zappasodi, P.~M. Rossini, Mapping
  distributed sources of cortical rhythms in mild {{Alzheimer}}'s disease.
  {{A}} multicentric {{EEG}} study, Neuroimage 22~(1) (2004) 57--67.
\newblock \href {http://dx.doi.org/10.1016/j.neuroimage.2003.09.028}
  {\path{doi:10.1016/j.neuroimage.2003.09.028}}.

\bibitem{jeong_eeg_2004}
J.~Jeong, {{EEG}} dynamics in patients with {{Alzheimer}}'s disease, Clin
  Neurophysiol 115~(7) (2004) 1490--1505.
\newblock \href {http://dx.doi.org/10.1016/j.clinph.2004.01.001}
  {\path{doi:10.1016/j.clinph.2004.01.001}}.

\bibitem{dauwels_diagnosis_2010}
J.~Dauwels, F.~Vialatte, A.~Cichocki, Diagnosis of {{Alzheimer}}'s {{Disease}}
  from {{EEG Signals}}: {{Where Are We Standing}}?, Current Alzheimer Research
  7~(6) (2010) 487--505.
\newblock \href {http://dx.doi.org/10.2174/156720510792231720}
  {\path{doi:10.2174/156720510792231720}}.

\bibitem{wang_power_2015}
R.~Wang, J.~Wang, H.~Yu, X.~Wei, C.~Yang, B.~Deng,
  \href{http://www.ncbi.nlm.nih.gov/pmc/articles/PMC4427585/}{Power spectral
  density and coherence analysis of {{Alzheimer}}'s {{EEG}}}, Cogn Neurodyn
  9~(3) (2015) 291--304.
\newblock \href {http://dx.doi.org/10.1007/s11571-014-9325-x}
  {\path{doi:10.1007/s11571-014-9325-x}}.

\bibitem{buckner_brains_2008}
R.~L. Buckner, J.~R. Andrews-Hanna, D.~L. Schacter,
  \href{http://onlinelibrary.wiley.com/doi/10.1196/annals.1440.011/abstract}{The
  {{Brain}}'s {{Default Network}}}, Annals of the New York Academy of Sciences
  1124~(1) (2008) 1--38.
\newblock \href {http://dx.doi.org/10.1196/annals.1440.011}
  {\path{doi:10.1196/annals.1440.011}}.

\bibitem{stam_magnetoencephalographic_2006}
C.~J. Stam, B.~F. Jones, I.~Manshanden, A.~M. {van Cappellen van Walsum},
  T.~Montez, J.~P.~A. Verbunt, J.~C. {de Munck}, B.~W. {van Dijk}, H.~W.
  Berendse, P.~Scheltens,
  \href{http://www.sciencedirect.com/science/article/pii/S105381190600601X}{Magnetoencephalographic
  evaluation of resting-state functional connectivity in {{Alzheimer}}'s
  disease}, NeuroImage 32~(3) (2006) 1335--1344.
\newblock \href {http://dx.doi.org/10.1016/j.neuroimage.2006.05.033}
  {\path{doi:10.1016/j.neuroimage.2006.05.033}}.

\bibitem{purves_neuroscience_2001}
D.~Purves, G.~J. Augustine, D.~Fitzpatrick, L.~C. Katz, A.-S. LaMantia, J.~O.
  McNamara, S.~M. Williams (Eds.), Neuroscience, 2nd Edition, {Sinauer
  Associates}, 2001.

\bibitem{pearson_anatomical_1985}
R.~C. Pearson, M.~M. Esiri, R.~W. Hiorns, G.~K. Wilcock, T.~P. Powell,
  Anatomical correlates of the distribution of the pathological changes in the
  neocortex in {{Alzheimer}} disease, Proc. Natl. Acad. Sci. U.S.A. 82~(13)
  (1985) 4531--4534.

\bibitem{arnold_topographical_1991}
S.~E. Arnold, B.~T. Hyman, J.~Flory, A.~R. Damasio, G.~W. Van~Hoesen, The
  topographical and neuroanatomical distribution of neurofibrillary tangles and
  neuritic plaques in the cerebral cortex of patients with {{Alzheimer}}'s
  disease, Cereb. Cortex 1~(1) (1991) 103--116.

\bibitem{catani_rises_2005}
M.~Catani, D.~H. Ffytche, The rises and falls of disconnection syndromes, Brain
  128~(Pt 10) (2005) 2224--2239.
\newblock \href {http://dx.doi.org/10.1093/brain/awh622}
  {\path{doi:10.1093/brain/awh622}}.

\bibitem{miltner_coherence_1999-1}
W.~H. Miltner, C.~Braun, M.~Arnold, H.~Witte, E.~Taub, Coherence of gamma-band
  {{EEG}} activity as a basis for associative learning, Nature 397~(6718)
  (1999) 434--436.
\newblock \href {http://dx.doi.org/10.1038/17126} {\path{doi:10.1038/17126}}.

\bibitem{buschman_top-down_2007}
T.~J. Buschman, E.~K. Miller, Top-down versus bottom-up control of attention in
  the prefrontal and posterior parietal cortices, Science 315~(5820) (2007)
  1860--1862.
\newblock \href {http://dx.doi.org/10.1126/science.1138071}
  {\path{doi:10.1126/science.1138071}}.

\bibitem{siegel_neuronal_2008}
M.~Siegel, T.~H. Donner, R.~Oostenveld, P.~Fries, A.~K. Engel,
  \href{http://www.sciencedirect.com/science/article/pii/S0896627308007575}{Neuronal
  {{Synchronization}} along the {{Dorsal Visual Pathway Reflects}} the
  {{Focus}} of {{Spatial Attention}}}, Neuron 60~(4) (2008) 709--719.
\newblock \href {http://dx.doi.org/10.1016/j.neuron.2008.09.010}
  {\path{doi:10.1016/j.neuron.2008.09.010}}.

\bibitem{gregoriou_high-frequency_2009}
G.~G. Gregoriou, S.~J. Gotts, H.~Zhou, R.~Desimone, High-frequency, long-range
  coupling between prefrontal and visual cortex during attention, Science
  324~(5931) (2009) 1207--1210.
\newblock \href {http://dx.doi.org/10.1126/science.1171402}
  {\path{doi:10.1126/science.1171402}}.

\bibitem{hipp_oscillatory_2011}
J.~F. Hipp, A.~K. Engel, M.~Siegel, Oscillatory synchronization in large-scale
  cortical networks predicts perception, Neuron 69~(2) (2011) 387--396.
\newblock \href {http://dx.doi.org/10.1016/j.neuron.2010.12.027}
  {\path{doi:10.1016/j.neuron.2010.12.027}}.

\bibitem{canolty_high_2006}
R.~T. Canolty, E.~Edwards, S.~S. Dalal, M.~Soltani, S.~S. Nagarajan, H.~E.
  Kirsch, M.~S. Berger, N.~M. Barbaro, R.~T. Knight,
  \href{http://www.ncbi.nlm.nih.gov/pmc/articles/PMC2628289/}{High {{Gamma
  Power Is Phase}}-{{Locked}} to {{Theta Oscillations}} in {{Human
  Neocortex}}}, Science 313~(5793) (2006) 1626--1628.
\newblock \href {http://dx.doi.org/10.1126/science.1128115}
  {\path{doi:10.1126/science.1128115}}.

\bibitem{axmacher_cross-frequency_2010}
N.~Axmacher, M.~M. Henseler, O.~Jensen, I.~Weinreich, C.~E. Elger, J.~Fell,
  \href{http://www.pnas.org/content/107/7/3228}{Cross-frequency coupling
  supports multi-item working memory in the human hippocampus}, PNAS 107~(7)
  (2010) 3228--3233.
\newblock \href {http://dx.doi.org/10.1073/pnas.0911531107}
  {\path{doi:10.1073/pnas.0911531107}}.

\bibitem{goutagny_alterations_2013}
R.~Goutagny, N.~Gu, C.~Cavanagh, J.~Jackson, J.-G. Chabot, R.~Quirion,
  S.~Krantic, S.~Williams,
  \href{http://onlinelibrary.wiley.com/doi/10.1111/ejn.12233/abstract}{Alterations
  in hippocampal network oscillations and theta\textendash{}gamma coupling
  arise before {{A$\beta$}} overproduction in a mouse model of {{Alzheimer}}'s
  disease}, Eur J Neurosci 37~(12) (2013) 1896--1902.
\newblock \href {http://dx.doi.org/10.1111/ejn.12233}
  {\path{doi:10.1111/ejn.12233}}.

\bibitem{li_discriminant_2012}
Y.~Li, Y.~Wang, G.~Wu, F.~Shi, L.~Zhou, W.~Lin, D.~Shen, {Alzheimer's Disease
  Neuroimaging Initiative}, Discriminant analysis of longitudinal cortical
  thickness changes in {{Alzheimer}}'s disease using dynamic and network
  features, Neurobiol. Aging 33~(2) (2012) 427.e15--30.
\newblock \href {http://dx.doi.org/10.1016/j.neurobiolaging.2010.11.008}
  {\path{doi:10.1016/j.neurobiolaging.2010.11.008}}.

\bibitem{wang_disrupted_2013}
J.~Wang, X.~Zuo, Z.~Dai, M.~Xia, Z.~Zhao, X.~Zhao, J.~Jia, Y.~Han, Y.~He,
  Disrupted functional brain connectome in individuals at risk for
  {{Alzheimer}}'s disease, Biol. Psychiatry 73~(5) (2013) 472--481.
\newblock \href {http://dx.doi.org/10.1016/j.biopsych.2012.03.026}
  {\path{doi:10.1016/j.biopsych.2012.03.026}}.

\bibitem{wee_enriched_2011}
C.-Y. Wee, P.-T. Yap, W.~Li, K.~Denny, J.~N. Browndyke, G.~G. Potter, K.~A.
  Welsh-Bohmer, L.~Wang, D.~Shen, Enriched white matter connectivity networks
  for accurate identification of {{MCI}} patients, Neuroimage 54~(3) (2011)
  1812--1822.
\newblock \href {http://dx.doi.org/10.1016/j.neuroimage.2010.10.026}
  {\path{doi:10.1016/j.neuroimage.2010.10.026}}.

\bibitem{wee_identification_2012}
C.-Y. Wee, P.-T. Yap, D.~Zhang, K.~Denny, J.~N. Browndyke, G.~G. Potter, K.~A.
  Welsh-Bohmer, L.~Wang, D.~Shen, Identification of {{MCI}} individuals using
  structural and functional connectivity networks, Neuroimage 59~(3) (2012)
  2045--2056.
\newblock \href {http://dx.doi.org/10.1016/j.neuroimage.2011.10.015}
  {\path{doi:10.1016/j.neuroimage.2011.10.015}}.

\bibitem{horwitz_functional_2011}
B.~Horwitz, J.~B. Rowe,
  \href{http://www.sciencedirect.com/science/article/pii/S030100821100116X}{Functional
  biomarkers for neurodegenerative disorders based on the network paradigm},
  Progress in Neurobiology 95~(4) (2011) 505--509.
\newblock \href {http://dx.doi.org/10.1016/j.pneurobio.2011.07.005}
  {\path{doi:10.1016/j.pneurobio.2011.07.005}}.

\bibitem{hutchison_network-based_2011}
D.~Dai, H.~He, J.~Vogelstein, Z.~Hou,
  \href{http://link.springer.com/10.1007/978-3-642-24319-6_24}{Network-{{Based
  Classification Using Cortical Thickness}} of {{AD Patients}}}, in:
  D.~Hutchison, T.~Kanade, J.~Kittler, J.~M. Kleinberg, F.~Mattern, J.~C.
  Mitchell, M.~Naor, O.~Nierstrasz, C.~Pandu~Rangan, B.~Steffen, M.~Sudan,
  D.~Terzopoulos, D.~Tygar, M.~Y. Vardi, G.~Weikum, K.~Suzuki, F.~Wang,
  D.~Shen, P.~Yan (Eds.), Machine {{Learning}} in {{Medical Imaging}}, Vol.
  7009, {Springer Berlin Heidelberg}, Berlin, Heidelberg, 2011, pp. 193--200.

\bibitem{shao_prediction_2012}
J.~Shao, N.~Myers, Q.~Yang, J.~Feng, C.~Plant, C.~B{\"o}hm, H.~F{\"o}rstl,
  A.~Kurz, C.~Zimmer, C.~Meng, V.~Riedl, A.~Wohlschl{\"a}ger, C.~Sorg,
  \href{http://www.ncbi.nlm.nih.gov/pmc/articles/PMC3778749/}{Prediction of
  {{Alzheimer}}'s disease using individual structural connectivity networks},
  Neurobiol Aging 33~(12) (2012) 2756--2765.
\newblock \href {http://dx.doi.org/10.1016/j.neurobiolaging.2012.01.017}
  {\path{doi:10.1016/j.neurobiolaging.2012.01.017}}.

\bibitem{zhou_hierarchical_2011}
L.~Zhou, Y.~Wang, Y.~Li, P.-T. Yap, D.~Shen, (adni), the Alzheimer's Disease
  Neuroimaging~Initiative,
  \href{http://journals.plos.org/plosone/article?id=10.1371/journal.pone.0021935}{Hierarchical
  {{Anatomical Brain Networks}} for {{MCI Prediction}}: {{Revisiting Volumetric
  Measures}}}, PLOS ONE 6~(7) (2011) e21935.
\newblock \href {http://dx.doi.org/10.1371/journal.pone.0021935}
  {\path{doi:10.1371/journal.pone.0021935}}.

\bibitem{dai_discriminative_2012}
Z.~Dai, C.~Yan, Z.~Wang, J.~Wang, M.~Xia, K.~Li, Y.~He,
  \href{http://www.sciencedirect.com/science/article/pii/S1053811911011645}{Discriminative
  analysis of early {{Alzheimer}}'s disease using multi-modal imaging and
  multi-level characterization with multi-classifier ({{M3}})}, NeuroImage
  59~(3) (2012) 2187--2195.
\newblock \href {http://dx.doi.org/10.1016/j.neuroimage.2011.10.003}
  {\path{doi:10.1016/j.neuroimage.2011.10.003}}.

\bibitem{shu_disrupted_2012}
N.~Shu, Y.~Liang, H.~Li, J.~Zhang, X.~Li, L.~Wang, Y.~He, Y.~Wang, Z.~Zhang,
  Disrupted topological organization in white matter structural networks in
  amnestic mild cognitive impairment: Relationship to subtype, Radiology
  265~(2) (2012) 518--527.
\newblock \href {http://dx.doi.org/10.1148/radiol.12112361}
  {\path{doi:10.1148/radiol.12112361}}.

\bibitem{stam_small-world_2007}
C.~J. Stam, B.~F. Jones, G.~Nolte, M.~Breakspear, P.~Scheltens, Small-world
  networks and functional connectivity in {{Alzheimer}}'s disease, Cereb.
  Cortex 17~(1) (2007) 92--99.
\newblock \href {http://dx.doi.org/10.1093/cercor/bhj127}
  {\path{doi:10.1093/cercor/bhj127}}.

\bibitem{grober_free_2010}
E.~Grober, A.~E. Sanders, C.~Hall, R.~B. Lipton, Free and cued selective
  reminding identifies very mild dementia in primary care, Alzheimer Dis Assoc
  Disord 24~(3) (2010 Jul-Sep) 284--290.
\newblock \href {http://dx.doi.org/10.1097/WAD.0b013e3181cfc78b}
  {\path{doi:10.1097/WAD.0b013e3181cfc78b}}.

\bibitem{velayudhan_review_2014}
L.~Velayudhan, S.-H. Ryu, M.~Raczek, M.~Philpot, J.~Lindesay, M.~Critchfield,
  G.~Livingston,
  \href{http://www.ncbi.nlm.nih.gov/pmc/articles/PMC4071993/}{Review of brief
  cognitive tests for patients with suspected dementia}, Int Psychogeriatr
  26~(8) (2014) 1247--1262.
\newblock \href {http://dx.doi.org/10.1017/S1041610214000416}
  {\path{doi:10.1017/S1041610214000416}}.

\bibitem{tombaugh_mini-mental_1992}
T.~N. Tombaugh, N.~J. McIntyre, The mini-mental state examination: A
  comprehensive review, J Am Geriatr Soc 40~(9) (1992) 922--935.

\bibitem{sperling_functional_2010}
R.~A. Sperling, B.~C. Dickerson, M.~Pihlajamaki, P.~Vannini, P.~S. LaViolette,
  O.~V. Vitolo, T.~Hedden, J.~A. Becker, D.~M. Rentz, D.~J. Selkoe, K.~A.
  Johnson,
  \href{http://www.ncbi.nlm.nih.gov/pmc/articles/PMC3036844/}{Functional
  {{Alterations}} in {{Memory Networks}} in {{Early Alzheimer}}'s {{Disease}}},
  Neuromolecular Med 12~(1) (2010) 27--43.
\newblock \href {http://dx.doi.org/10.1007/s12017-009-8109-7}
  {\path{doi:10.1007/s12017-009-8109-7}}.

\bibitem{buckner_molecular_2005}
R.~L. Buckner, A.~Z. Snyder, B.~J. Shannon, G.~LaRossa, R.~Sachs, A.~F.
  Fotenos, Y.~I. Sheline, W.~E. Klunk, C.~A. Mathis, J.~C. Morris, M.~A.
  Mintun, Molecular, structural, and functional characterization of
  {{Alzheimer}}'s disease: Evidence for a relationship between default
  activity, amyloid, and memory, J. Neurosci. 25~(34) (2005) 7709--7717.
\newblock \href {http://dx.doi.org/10.1523/JNEUROSCI.2177-05.2005}
  {\path{doi:10.1523/JNEUROSCI.2177-05.2005}}.

\bibitem{greicius_default-mode_2004}
M.~D. Greicius, G.~Srivastava, A.~L. Reiss, V.~Menon,
  \href{http://www.ncbi.nlm.nih.gov/pmc/articles/PMC384799/}{Default-mode
  network activity distinguishes {{Alzheimer}}'s disease from healthy aging:
  {{Evidence}} from functional {{MRI}}}, Proc Natl Acad Sci U S A 101~(13)
  (2004) 4637--4642.
\newblock \href {http://dx.doi.org/10.1073/pnas.0308627101}
  {\path{doi:10.1073/pnas.0308627101}}.

\bibitem{srinivasan_eeg_2007}
R.~Srinivasan, W.~R. Winter, J.~Ding, P.~L. Nunez, {{EEG}} and {{MEG}}
  coherence: Measures of functional connectivity at distinct spatial scales of
  neocortical dynamics, J. Neurosci. Methods 166~(1) (2007) 41--52.
\newblock \href {http://dx.doi.org/10.1016/j.jneumeth.2007.06.026}
  {\path{doi:10.1016/j.jneumeth.2007.06.026}}.

\bibitem{schoffelen_source_2009}
J.-M. Schoffelen, J.~Gross, Source connectivity analysis with {{MEG}} and
  {{EEG}}, Hum Brain Mapp 30~(6) (2009) 1857--1865.
\newblock \href {http://dx.doi.org/10.1002/hbm.20745}
  {\path{doi:10.1002/hbm.20745}}.

\bibitem{colclough_how_2016}
G.~L. Colclough, M.~W. Woolrich, P.~K. Tewarie, M.~J. Brookes, A.~J. Quinn,
  S.~M. Smith, How reliable are {{MEG}} resting-state connectivity metrics?,
  Neuroimage 138 (2016) 284--293.
\newblock \href {http://dx.doi.org/10.1016/j.neuroimage.2016.05.070}
  {\path{doi:10.1016/j.neuroimage.2016.05.070}}.

\bibitem{nolte_identifying_2004}
G.~Nolte, O.~Bai, L.~Wheaton, Z.~Mari, S.~Vorbach, M.~Hallett,
  \href{http://www.clinph-journal.com/article/S1388245704001993/abstract}{Identifying
  true brain interaction from {{EEG}} data using the imaginary part of
  coherency}, Clinical Neurophysiology 115~(10) (2004) 2292--2307.
\newblock \href {http://dx.doi.org/10.1016/j.clinph.2004.04.029}
  {\path{doi:10.1016/j.clinph.2004.04.029}}.

\bibitem{boccaletti_structure_2014}
S.~Boccaletti, G.~Bianconi, R.~Criado, C.~I. {del Genio},
  J.~G{\'o}mez-Garde{\~n}es, M.~Romance, I.~Sendi{\~n}a-Nadal, Z.~Wang,
  M.~Zanin,
  \href{http://www.sciencedirect.com/science/article/pii/S0370157314002105}{The
  structure and dynamics of multilayer networks}, Physics Reports 544~(1)
  (2014) 1--122.
\newblock \href {http://dx.doi.org/10.1016/j.physrep.2014.07.001}
  {\path{doi:10.1016/j.physrep.2014.07.001}}.

\end{thebibliography}
